\documentclass{aa}  

\usepackage{graphicx}
\usepackage{txfonts}
\usepackage{natbib}
\usepackage{threeparttable}
\usepackage{supertabular}
\usepackage{booktabs}
\usepackage{array}
\begin{document}

   \subtitle{Results from the CALYPSO IRAM-PdBI survey}

   \title{Probing the CO and methanol snow lines in young protostars\thanks{Based on observations carried out under project number U052 with the IRAM Plateau de Bure Interferometer. IRAM is supported by INSU/CNRS (France), MPG (Germany) and IGN (Spain).}}
  \titlerunning{Probing the CO and methanol snow lines in young protostars}
  
   \author{S. Anderl\inst{1,2,}, 
          S. Maret\inst{1,2},
          S. Cabrit\inst{3,1,2},
          A. Belloche\inst{4},
          A.~J. Maury\inst{5},
          Ph. Andr{\'e}\inst{5},
          C. Codella\inst{6},
          A. Bacmann\inst{1,2},
          S. Bontemps\inst{7,8},
          L. Podio\inst{6},
         F. Gueth\inst{9},
         E. Bergin\inst{10}
                  }
      \authorrunning{S.Anderl et al.}

   \institute{ Univ. Grenoble Alpes, IPAG, F-38000 Grenoble, France\\
    \email{sibylle.anderl@obs.ujf-grenoble.fr}
   \and
CNRS, IPAG, F-38000 Grenoble, France
\and
LERMA, Observatoire de Paris, PSL Research University, CNR
S, UMR 8112, F-75014, Paris France
\and
Max-Planck-Institut f{\"u}r Radioastronomie, Auf dem H{\"u}gel 69, 53121 Bonn, Germany
\and
Laboratoire AIM-Paris-Saclay, CEA/DSM/Irfu - CNRS - Universit{\'e} Paris Diderot, CE-Saclay, 91191 Gif-sur-Yvette, France
\and
INAF - Osservatorio Astrofisico di Arcetri, Largo E. Fermi 5, 50125 Firenze, Italy
\and
Univ. Bordeaux, LAB, UMR 5804, F-33270, Floirac, France       
\and
CNRS, LAB, UMR 5804, F-33270 Floirac, France
\and
IRAM, 300 rue de la piscine, 38406 Saint Martin d'H{\`e}res, France
\and
Department of Astronomy, University of Michigan, 500 Church St., Ann Arbor, MI 48109, USA
                                }

   \date{Received ...; accepted ...}

 
  \abstract
%
   {``Snow lines'', marking regions where abundant volatiles freeze out onto the surface of dust grains, play an important role for planet growth and bulk composition in protoplanetary disks. They can already be observed in the envelopes of the much younger, low-mass Class 0 protostars that are still in their early phase of heavy accretion.}
   {We aim at using the information on the sublimation regions of different kinds of ices to understand the chemistry of the envelope, its temperature and density structure, and the history of the accretion process. 
}
   {As part of the CALYPSO IRAM Large Program, we have obtained observations of C$^{18}$O, N$_2$H$^+$ and CH$_3$OH towards nearby Class 0 protostars with the IRAM Plateau de Bure interferometer at sub-arcsecond resolution. For four of these sources we have modeled the emission using a chemical code coupled with a radiative transfer module.
   }
   {We observe an anti-correlation of C$^{18}$O and N$_2$H$^+$ in NGC 1333-IRAS4A, NGC 1333-IRAS4B, L1157, and L1448C, with N$_2$H$^+$ forming a ring around the centrally peaked C$^{18}$O emission due to N$_2$H$^+$ being chemically destroyed by CO. The emission regions of models and observations match for a CO binding energy of 1200 K, which is higher than the binding energy of pure CO ices ($\sim$855 K). Furthermore, we find very low CO abundances inside the snow lines in our sources, about an order of magnitude lower than the total CO abundance observed in the gas on large scales in molecular clouds before depletion sets in.}
   {The high CO binding energy may hint at CO being frozen out in a polar ice environment like amorphous water ice or in non-polar CO$_2$-rich ice. The low CO abundances are comparable to values found in protoplanetary disks, which may indicate an evolutionary scenario where these low values are already established in the protostellar phase.}


   \maketitle
%

\section{Introduction }

The earliest phases of star formation take place deeply embedded in dusty molecular clouds. Most of the mass of the youngest low-mass, so-called Class 0 protostars (\citealt{Andre:1993}) is still in the collapsing envelope, and their spectral energy distribution is entirely dominated by the emission of heated dust in the envelope. A characteristic feature of these very young protostars is the existence of well-collimated outflows, which have been one of the main routes to discovering these highly obscured stars. The Class 0 phase is crucial for the subsequent evolution of sun-like stars as it sets the initial conditions for the establishment of their final properties, like their final mass or their protoplanetary disk. However, their structure on scales of 100 au is still poorly known and many open questions, e.g. regarding their multiplicity, the launching mechanism of their outflows, the existence of protostellar disks, and the physical and chemical structure of their envelopes, still remain. An understanding of the chemistry around young protostars, in particular, is crucial in order to understand the composition of the material that eventually forms planets. Furthermore, it has been suggested to use the composition of the envelope as a chemical clock that can be used to understand the evolution of these young stars after they started heating up their environment (e.g. \citealt{Jorgensen:2002}). The chemical structure of the envelope may also trace the history of the accretion process, since a sudden increase in accretion luminosity of the protostar will temporarily change the temperature structure. This change will still be visible in the chemistry of the envelope even when the accretion burst is over, because the chemistry does not instantaneously adapt to the changes (e.g. \citealt{Jorgensen:2015}).

Protostellar envelopes have large density and temperature gradients. These gradients cause a chemical stratification, with the most volatile of the species that are found in ices, such as CO, being present as vapor in rather large parts of the envelope, while the less volatile components are present in the gas phase only in the innermost region. More specifically, each species that is depleted in the parts of the envelope where the temperature is low and densities are high enough for depletion to occur has a characteristic freeze-out behaviour. This behaviour, in simplified terms, results in a chemical step profile, where the chemical abundance jumps from the low, depleted value to a higher abundance in the inner envelope at a radius where the dust temperature corresponds to the freeze-out temperature of the respective species (e.g. \citealt{Maret:2002}). For instance, the freeze-out temperature of pure CO ice lies between 20 and 30 K, while pure methanol ice is believed to sublimate at higher temperatures of about 90-120 K and is hence only expected in the close proximity of the protostar (e.g.  \citealt{Sandford:1993,Collings:2004}). In protoplanetary disks, the radii where abundant volatiles freeze out of the gas phase onto the grains are termed ``snow lines''. In the following, we adopt this term for protostellar environments.

The fact that CO is depleted in the outer envelope can also be seen from the existence of species in the gas-phase that are chemically anti-correlated with CO. A prominent example of such a species is N$_2$H$^+$, as N$_2$H$^+$ gets chemically destroyed by CO (e.g. \citealt{Bergin:2001,Maret:2006}). Accordingly, the spatially resolved observation of anti-correlated emission of CO and N$_2$H$^+$ is a particularly reliable tracer for the transition from CO depletion to a high gas-phase abundance of CO, while an analysis of the intensity of CO emission alone may be affected by excitation and optical depth effects (\citealt{Qi:2013,Qi:2015}).

In this study, we aim at probing the CO and methanol snow lines in young protostars and use this information to draw conclusions about the chemical and physical structure of the envelope, the physics of ice depletion and sublimation, and the history of the accretion process. This paper is the fifth in a series of publications resulting from the CALYPSO (Continuum And Lines in Young Proto-Stellar Objects\footnote{\url{http://irfu.cea.fr/Projets/Calypso/Welcome.html}}) survey. This survey utilises the Plateau de Bure interferometer (PdBI) to conduct a large observational program that aims at studying a large sample of Class 0 protostars at sub-arcsecond resolution. Nearby (d $<$ 300 pc) Class 0 protostars observable with the PdBI were mapped in dust continuum at 1 and 3mm, together with high spectral resolution observations of molecular tracers of the outflow, the inner and the outer envelope, and broad-band spectra in all bands (see also \citealt{Maret:2014,Maury:2014}, and \citealt{Codella:2014} for initial CALYPSO results). In this paper we analyse high spatial and spectral resolution observations of C$^{18}$O, N$_2$H$^+$, and CH$_3$OH in four of the observed sources. While our angular resolution is clearly high enough to spatially resolve the CO snow line, we note that we do not necessarily expect the methanol emission to  be resolved by our observations, because the temperature for methanol desorption will only be achieved very close to the central source.
We restrict this study to the sources in our sample that show a clear anti-correlation of C$^{18}$O and N$_2$H$^+$ in order to unambiguously trace the radius where CO gets depleted on grain surfaces (see above).

This paper is structured as follows: In Section \ref{Observations} our observations are presented. In Section \ref{Results} we first give an overview of the four sources that show an anti-correlation of C$^{18}$O and N$_2$H$^+$ (Section \ref{thesources}), then discuss their emission morphologies in Section \ref{snowlines}, and finally present the emission size measurements from uv-fits of the observations (Section \ref{snowlines1}). In Section \ref{analysis} we  present our modelling approach (\ref{model}), its application on IRAS4B as a testcase (Section \ref{testcase}), and on the other sources (Section \ref{othersources}). We then discuss our results in Section \ref{Discussion} and summarise our findings in Section \ref{Conclusions}.

\section{Observations}\label{Observations}

Observations towards the four sources (see Table \ref{sources}) were performed with the PdBI between November 2010 and November 2011 using the A and C configurations of the array. All lines were observed using the narrow-band backend. For the C$^{18}$O (2--1) line at 219.560 350 GHz (1.37 mm), this backend provided a bandwidth of 250 channels of 73 kHz (0.1 km s$^{-1}$) each, while for the N$_2$H$^+$ (1--0) hyperfine lines centered at 93.173 800 GHz (3.22 mm), the bandwidth corresponds to 512 channels of 39 kHz (0.13 km s$^{-1}$). The CH$_3$OH (5$_1$--4$_2$ $\varv_{\rm t}$ = 0 E$_1$) line at 216.945 600 GHz (1.38 mm) was observed with 250 channels of 72 kHz (0.1 km s$^{-1}$) each. The N$_2$H$^+$ and CH$_3$OH data were then resampled at a spectral resolution of 0.15 and 0.2 km s$^{-1}$ (47 and 145 kHz), respectively. In the case of methanol this was to improve the signal-to-noise ratio. Calibration was done using \verb+CLIC+, which is part of the \verb+GILDAS+  software suite\footnote{\url{http://www.iram.fr/IRAMFR/GILDAS/}}. For the 1.4 mm observations, the phase RMS was $<$ 65$^{\circ}$, with precipitable water vapor (PWV) between 0.7 mm and 2.0 mm, and system temperatures (T$_{\rm sys}$) $<$ 150 K. All data with phase rms $>$ 50$^{\circ}$ was flagged before producing the uv-tables. For the 3 mm observations, the phase RMS was $<$ 50$^{\circ}$, with PWV between 0.9 mm and 3 mm and T$_{\rm sys}$ $<$ 80 K. Here, all data with phase rms $>$ 40$^{\circ}$ was flagged before producing the uv-tables. The continuum emission was removed from the visibility tables to produce continuum-free line tables. Spectral datacubes were produced from the visibility tables using a natural weighting, and deconvolved using the standard \verb+CLEAN+ algorithm in the \verb+MAPPING+ program. The synthesized beam sizes for all sources and lines, together with their RMS noise per channel in the final datacubes are listed in Table \ref{obs1}.

 \begin{table*}[!htb]
\caption{Sample of sources.}            
\label{sources}      
\centering                          
\begin{tabular}{l  l l l l l }    
\hline\hline
\noalign{\smallskip}
Source   & $\alpha_{\rm J2000}$\tablefoottext{a} &  $\delta_{\rm J2000}$\tablefoottext{a} & D\tablefoottext{b} & L$_{\rm bol}$\tablefoottext{b}& $\varv_{\rm LSR}$\tablefoottext{c}\\
& (h:m:s) &  ($^{\circ}$:$'$:$''$) & (pc) & (L$_{\sun}$)& (km s$^{-1}$)\\
\noalign{\smallskip}
\hline
\noalign{\smallskip}
IRAS4A\tablefoottext{d} & 03:29:10.43 & 31:13:32.2 & 235 &9.1 & +7.2\\
L1448C & 03:25:38.88 &  30:44:05.3 & 235 & 9.0 & +5.2\\
L1157 & 20:39:06.27  & 68:02:15.7& 325 &4.7 & +2.8\\
IRAS4B & 03:29:12.01 & 31:13:08.1& 235 &4.4 & +6.7\\
\noalign{\smallskip}
 \hline           
\end{tabular}
\tablefoot{
\tablefoottext{a}{Source positions were derived from uv-fits of the CALYPSO continuum data, see Maury et al. in prep.}
\tablefoottext{b}{Distances and bolometric luminosity estimates as used in Kristensen et al. (2012).}
\tablefoottext{c}{Obtained from C$^{18}$O spectra, extracted at the source positions. The systemic velocity is derived from Gaussian line fits or is assumed as the velocity where self-absorption is observed.}
\tablefoottext{d}{The coordinates correspond to the IRAS4A2 source. We detect compact methanol emission only towards this component of the IRAS4A binary system  (cf. \citealt{Santangelo:2015}).}
}
\end{table*}

 \begin{table*}[!htb]
\caption{Synthesized beam sizes and RMS noise per channel in the final datacubes for the observed lines and sources.}            
\label{obs1}      
\centering                          
\begin{tabular}{l  l l l l }     
\hline\hline
\noalign{\smallskip}
Source   & Line &  Synthesized half-power beam width\tablefoottext{a} & $\Delta \varv$& Noise per channel\tablefoottext{b}\\
& &  Major $\times$ Minor (Pos. Ang.) & (km s$^{-1}$)&(mJy/beam)\\
\noalign{\smallskip}
 \hline           
\noalign{\smallskip}
IRAS4A & C$^{18}$O (2--1) & 1.05$''$ $\times$ 0.79$''$ (199$^\circ$) & 0.10&21.9\\
&e-CH$_3$OH (5$_1$--4$_2$) & 1.10$''$ $\times$ 0.82$''$ (201$^\circ$) & 0.20& 16.6\\
& N$_2$H$^+$ (1--0)& 1.80$''$ $\times$ 1.32$''$ (38$^\circ$)&0.15&  5.8\\
L1448C & C$^{18}$O (2--1) &  1.01$''$ $\times$ 0.73$''$ (24$^\circ$) & 0.10 &  18.9\\
&e-CH$_3$OH (5$_1$--4$_2$) & 1.05$''$ $\times$ 0.74$''$ (26$^\circ$) & 0.20 &  13.8\\
& N$_2$H$^+$ (1--0)&  1.75$''$ $\times$ 1.40$''$ (64$^\circ$) & 0.15 &  5.5\\
L1157 & C$^{18}$O (2--1)  & 0.84$''$ $\times$ 0.76$''$ (-178$^\circ$)& 0.10 &23.5\\
&e-CH$_3$OH (5$_1$--4$_2$) & 0.87$''$ $\times$ 0.77$''$ (4$^\circ$) & 0.20 &  17.7\\
& N$_2$H$^+$ (1--0)&  1.49$''$ $\times$ 1.10$''$ (59$^\circ$) & 0.15 &  6.2\\
IRAS4B & C$^{18}$O (2--1) & 1.05$''$ $\times$ 0.81$''$ (-160$^\circ$) & 0.10&  21.9\\
&e-CH$_3$OH (5$_1$--4$_2$) & 1.10$''$ $\times$ 0.84$''$ (-161$^\circ$)& 0.20&16.6\\
& N$_2$H$^+$ (1--0)& 1.77$''$ $\times$ 1.28$''$ (38$^\circ$)& 0.15& 5.8\\
\noalign{\smallskip}
\hline
\end{tabular}
\tablefoot{
\tablefoottext{a}{The imaging was performed with natural weighting.}
\tablefoottext{b}{The reported value corresponds to the median noise level of all channel maps of the cube, the noise level in each channel being derived from a Gaussian fit to the distribution of intensities within the map of this channel.}
}
\end{table*}

\section{Results}\label{Results}
\subsection{Source sample properties}\label{thesources}

Within the sample of 16 sources that were observed within the CALYPSO survey, we found four sources that show an anti-correlation in their emission of C$^{18}$O and N$_2$H$^+$: NGC 1333-IRAS4B, NGC 1333-IRAS4A, L1157, and L1448-C (see Fig \ref{maps}). The source IRAM 04191 also shows a ring of emission in N$_2$H$^+$ with an approximate radius of 10$''$, as was already observed by \citet{Belloche:2004}. In our PdBI maps, this ring is however not filled by emission in C$^{18}$O, which is only weakly detected at 4$\sigma$ right towards the continuum peak with an FWHM extent of 1.4$''$. Because of this unusual emission morphology, which hints at different mechanisms being in play than in the other sources, we decided to focus our present analysis only on the other four clear-cut cases. We will however briefly comment on how our model applies to IRAM 04191 in Section \ref{othersources}. We note that all these sources belong to the list of ``confirmed Class 0 protostars'', for which initial estimates of L$_{\rm bol}$ and M$_{\rm env}$ exist (\citealt{Andre:2000}).  The emission of the remaining sources is discussed in Appendix \ref{others}, where we classify these sources according to the morphology of 
their C$^{18}$O and N$_2$H$^+$ emission.

The NGC-1333 IRAS 4 source is located in the Perseus molecular cloud and was discovered by \citet{Jennings:1987} using IRAS far-IR observations. Within the NGC 1333 star formation region, it is located in the south-east (e.g. \citealt{Knee:2000}). Its distance is estimated as 235 pc (\citealt{Hirota:2008}). \citet{Sandell:1991} discovered that NGC-1333 IRAS 4 is a binary, using JCMT submillimeter observations to identify the sources NGC-1333 IRAS 4A (hereafter IRAS4A) and NGC-1333 IRAS 4B (hereafter IRAS4B), separated by $\sim$30$''$. 

{\bf IRAS4A} is in itself a binary source consisting of IRAS4A1 and IRAS4A2, which are separated by about 1.8$''$ (\citealt{Looney:2000}). Towards this system, infall motion as indicated by inverse P Cygni profiles has been observed in several molecules (e.g. \citealt{DiFrancesco:2001,Belloche:2006}). While IRAS4A1 is more than three times brighter than its companion in millimeter and centimeter continuum, only IRAS4A2 shows spectra of high molecular complexity (\citealt{Santangelo:2015}). The system exhibits extended, well collimated, jetlike molecular outflows. The jet originating from IRAS4A2 shows a symmetric bending with a position angle of approximately 45$^\circ$ on large scales and 0$^\circ$ on the small scales up to about 4$''$ from the source (e.g. \citealt{Liseau:1988,Blake:1995,Yildiz:2012,Santangelo:2015}). The estimates for the dynamical age of the outflow lie between 5900 and 16000 yr (\citealt{Knee:2000,Yildiz:2012}). IRAS4A was one of the early candidates of a hot corino source and shows emission of complex molecules like HCOOCH$_3$, CH$_3$CN, CH$_2$(OH)CHO, or CH$_3$OCH$_3$ towards the innermost region of its envelope towards IRAS4A2 (\citealt{Bottinelli:2004,Bottinelli:2008,Taquet:2015}). 
Its bolometric luminosity is 9.1 L$_{\sun}$ according to \citet{Karska:2013} and 7.0 L$_{\sun}$ according to \citet{Dunham:2013}. As IRAS4A is a multiple source in a cluster-forming region (NGC 1333), the bolometric luminosity is somewhat uncertain because it results from integration of the global spectral energy distribution, including data points with relatively low angular resolution at long far-infrared wavelengths (e.g. 160-250 $\mu$m).
An alternate estimate of the internal luminosity of the protostar can be obtained from the flux of the protostar 
at 70 $\mu$m (\citealt{Dunham:2008}), a wavelength at which {\it Herschel} Gould Belt survey observations 
(\citealt{Andre:2010}) provide data at $\sim 8\arcsec $ resolution. Using the {\it Herschel} photometry derived 
by \citet{Sadavoy:2014} and Eq. (2) of \citet{Dunham:2008} leads to an internal luminosity of 3.4 L$_{\sun}$ for IRAS4A. 
In the modeling analysis presented below, we adopt the bolometric luminosity estimate of \citet{Karska:2013} 
but discuss the effect of the luminosity uncertainty on our results in Sect. 5.1.
The envelope mass of IRAS4A amounts to 5.6 M$_{\sun}$ according to \citet{Kristensen:2012}.

{\bf IRAS4B}, being located $\sim$30$''$ south-east of IRAS4A, has a companion 11$''$ to the east, which was named IRAS 4B$'$ by \citet{DiFrancesco:2001}. IRAS4B itself, however, does not show signs of multiplicity. IRAS4B shows clearly separated outflow lobes in north-south direction (e.g. \citealt{Choi:2001,DiFrancesco:2001,Jorgensen:2007, Yildiz:2012}). Inverse P Cygni profiles have been observed in $^{13}$CO, H$_2$CO, and HDO line emission that can be interpreted as signs of infall motions in the envelope (\citealt{DiFrancesco:2001, Jorgensen:2007,Coutens:2013}). Just as IRAS4A, IRAS4B harbors a hot corino as several observations of complex molecules (hereafter COMs) towards this source suggest (\citealt{Sakai:2006,Bottinelli:2007,Jorgensen:2010a,Jorgensen:2010b,Sakai:2012}). 
Its bolometric luminosity, however, is lower than for IRAS4A with an estimated value of 4.4 L$_{\sun}$ according to 
\citet{Karska:2013} or 3.7 L$_{\sun}$ according to \citet{Dunham:2013}. Like in the case of IRAS4A, the uncertainty
is quite large. The internal luminosity of IRAS4B is estimated to be only 1.5 L$_{\sun}$ from the 70 $\mu$m flux 
derived by \citet{Sadavoy:2014} from {\it Herschel} Gould Belt survey observations.
\citet{Kristensen:2012} estimated an envelope mass of 3.0 M$_{\sun}$.

{\bf L1448C} (also named L1448-mm) is also located in the Perseus molecular cloud, 1$^\circ$ southwest of and at the same distance as the star formation region NGC 1333 (\citealt{Hirota:2011}). It first attracted attention because of its powerful, bipolar parsec-scale outflow with a high terminal radial velocity of $\sim$70 km s$^{-1}$ and a small dynamical timescale of $\sim$3500 yr (\citealt{Bachiller:1990}), showing symmetric pairs of CO molecular bullets. Shortly afterwards, the driving source was discovered in cm (\citealt{Curiel:1990}) and mm wavelengths (\citealt{Bachiller:1991}). In the infrared, two counterparts of L1448-C were observed: L 1448 C(N) (which corresponds to the mm source) and L 1448 C(S) (\citealt{Jorgensen:2006}) or L1448-mm A and B (\citealt{Tobin:2007}), separated by 8$''$. The P.A. of the molecular outflow as observed in CO was first determined as $\sim$159$^\circ$ (\citealt{Bachiller:1995}). \citet{Hirano:2010} observed the underlying jet in SiO and found kinks in its P.A.: In the innermost region, the red-shifted jet component has a P.A. of 160$^\circ$, while the blue-shifted component has a P.A. of 155$^\circ$. 
\citet{Kristensen:2012} estimate the bolometric luminosity of L1448C as 9.0 L$_{\sun}$  and its envelope mass as 
3.9 M$_{\sun}$. In this case the internal luminosity derived from the {\it Herschel} 70 $\mu$m flux is 7.8 L$_{\sun}$, very 
similar to the bolometric luminosity estimate.

{\bf L1157-MM} is a young Class 0 protostar in the L1157 dark cloud. The distance is rather uncertain and could be 250, 300 or 450 pc (\citealt{Kun:1998,Looney:2007}). L1157-mm is driving the prototype of a so-called chemically rich outflow. Its outflow is characterised by a series of blue- and redshifted bow shocks interpreted as the impact of the precessing jets with the slow ambient medium and/or the outflow cavities (e.g. \citealt{Bachiller:1997,Bachiller:2001,Codella:2010}). The mean P.A. of the CO outflow is reported as 161$^\circ$ (\citealt{Bachiller:2001}), and models suggest a jet precession cone of 15$^\circ$ opening angle (\citealt{Gueth:1996,Zhang:2000, Bachiller:2001}). The jet is detected only within 10$''$ from the protostar (\citealt{Tafalla:2015}) and its direction has been recently measured as part of the CALYPSO project (Podio et al. in prep.). 
The envelope mass and the luminosity of this protostar were estimated as 1.5 M$_{\sun}$ and 4.7 L$_{\sun}$, respectively, 
by \citet{Kristensen:2012}, assuming a distance of 325 pc. The internal luminosity derived from the 
{\it Herschel} 70 $\mu$m flux is 3.9 L$_{\sun}$. 
In order to be consistent with the work of \citet{Kristensen:2012}, we adopt their values.

\subsection{Emission morphology}\label{snowlines}

\begin{figure}[!htb]
   \centering
   \includegraphics[width=0.5\textwidth]{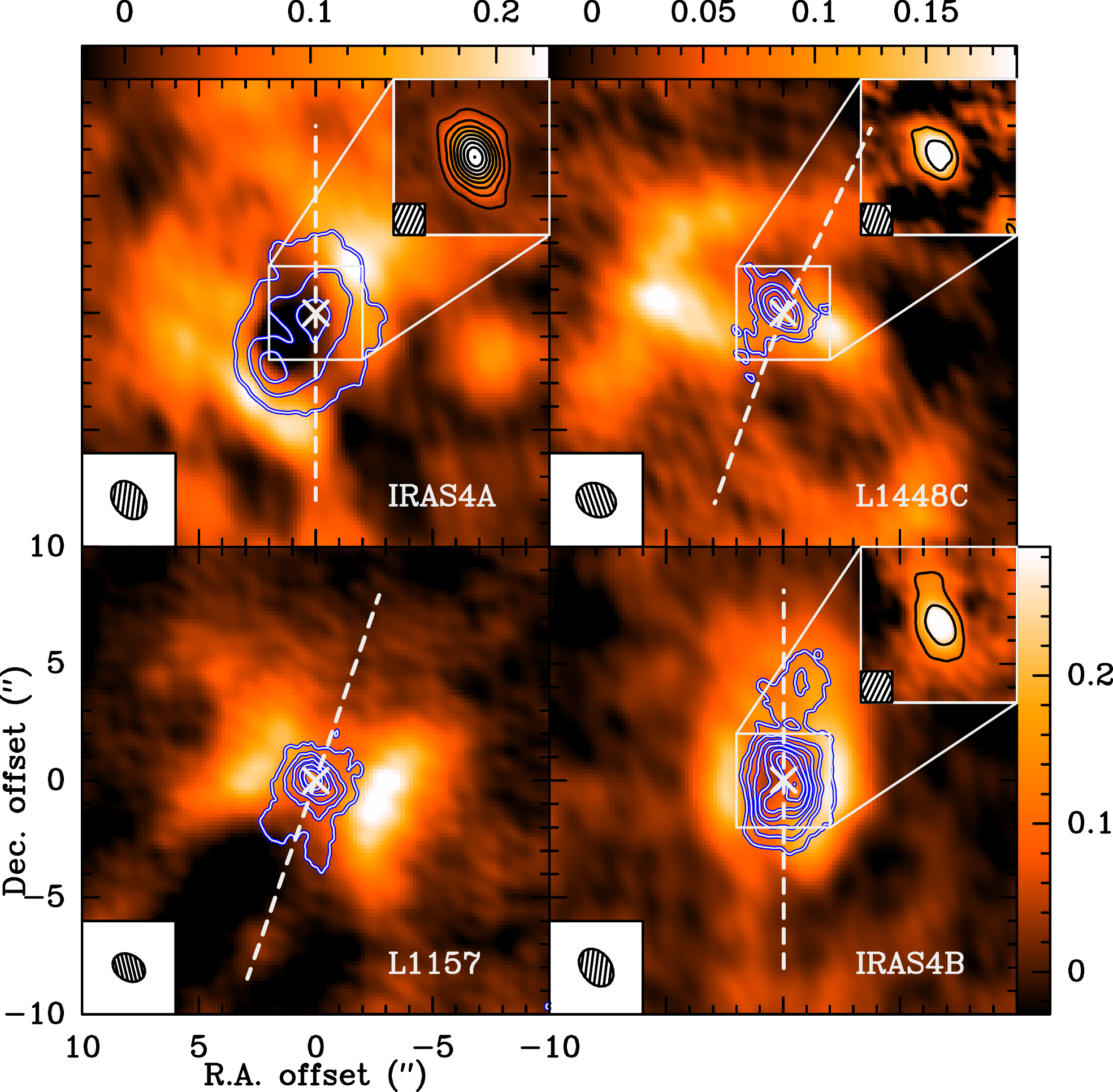}
      \caption{Anticorrelation between N$_2$H$^+$ and C$^{18}$O. Colour background: N$_2$H$^+$ (1--0) emission integrated over all seven hyperfine components.  The noise in these maps is $\sigma$=(0.030, 0.027, 0.025, 0.029) Jy beam$^{-1}$ km s$^{-1}$ for IRAS4A, L1448C, L1157, and IRAS4B, respectively. The wedges show the N$_2$H$^+$ intensity scale in Jy beam$^{-1}$ km s$^{-1}$. For the two lower panels the scaling is the same. Contours show emission  of C$^{18}$O (2--1) in steps of 6$\sigma$ (IRAS4A and 4B) or 3$\sigma$ (L1448C and L1157), starting at 3$\sigma$, with $\sigma$=(0.044, 0.029, 0.031, 0.033) Jy beam$^{-1}$ km s$^{-1}$ for IRAS4A, L1448C, L1157, and IRAS4B, respectively. The inlays in the upper right corners  show the methanol emission as colour background and contours in steps of 3$\sigma$, starting at 3$\sigma$ inside the central 2$''$, with  $\sigma$=(0.028, 0.019, 0.027) Jy beam$^{-1}$ km s$^{-1}$ for IRAS4A, L1448C, and IRAS4B. In L1157, the methanol line has not been detected at a 3$\sigma$ level. The filled ellipses in the lower left corner of the panels indicate the synthesized beam sizes of the N$_2$H$^+$ observations at 3 mm. The dashed white lines illustrate the small-scale outflow directions (see Table \ref{fitsc18o} and Section \ref{thesources}) and the white crosses show the positions of  the sources as listed in Table \ref{sources}. }
               \label{maps}
   \end{figure}

Figure \ref{maps} shows the observed integrated emission of C$^{18}$O, N$_2$H$^+$, and CH$_3$OH, where the lines of C$^{18}$O and CH$_3$OH were integrated over $\pm$ 3 km s$^{-1}$ around the systemic velocity of each source (see Table \ref{sources}), while the emission of N$_2$H$^+$ was integrated over a window of 20 km s$^{-1}$ that covers all the hyperfine components of the (1--0) transition. We derived the systemic velocities for the sources from inspection of the C$^{18}$O spectra extracted at the source position. This procedure might be prone to uncertainties because of the unknown dynamics of the inner envelope. Our analysis, however, does not depend on the exact values of the systemic velocities because we only use spectral information integrated over the full line width. We note that the intensity ratio of the N$_2$H$^+$ hyperfine lines, fitted with {\tt CLASS} and averaged within the central 10$''$ towards IRAS4B, IRAS4A, L1448C, and L1157, is 3:5:1.3, 3:4.6:1.4, 3:4.9:1.8, 3:5.1:1.1, respectively. These ratios suggest that the emission in the central regions is, on average, optically thin, where the ratio would be 3:5:1.

{\bf C$^{18}$O morphology:} In IRAS4B and L1157, the emission of C$^{18}$O is clearly elongated in the direction of the outflow, while for IRAS4A and L1448C the influence of the outflow on the C$^{18}$O is less clear. In the case of IRAS4A, the emission is elongated both along the inner outflow P.A. (north-south) and towards IRAS4A1 in the southwest. However, it still clearly peaks on A2, consistent with this COM source being the main heating source (cf. \citet{Santangelo:2015}).  
In L1448C, the emission region is elongated almost perpendicular to the outflow, unlike in the other sources. 

{\bf N$_2$H$^{+}$ morphology:} For all four sources, we clearly resolve the expected rings of N$_2$H$^+$ emission around the region where CO is sublimated. Peaks of N$_2$H$^+$ are found in regions roughly perpendicular to the outflow direction. These peaks closely adjoin the C$^{18}$O emission, strongly suggesting that the C$^{18}$O and N$_2$H$^+$ emission extents in a direction perpendicular to the outflow are dominated by CO and N$_2$ ice sublimation.
The N$_2$H$^+$ rings are perturbed along the outflow axis. The N$_2$H$^+$ abundance is indeed expected to decrease in shocks along the outflow due to electronic recombination (e.g. \citealt{Podio:2014}). An interesting case is IRAS4A, where the N$_2$H$^+$ emission seems to drop along the direction of the large-scale outflow at a P.A. of $\sim$45$^\circ$. The outflow direction changes about 10$''$ away from the central source (\citealt{Yildiz:2012}), which corresponds to a timescale of $\sim$600 yr. The observed N$_2$H$^+$ "hole" at P.A. =45$^\circ$ also coincides with the EHV peak in the A1 outflow seen by \citet{Santangelo:2015} (right panels in their Fig 5).
A somewhat puzzling case in terms of its emission in N$_2$H$^+$ is L1448C. Contrary to the other sources, the peaks in N$_2$H$^+$ emission are not equidistant to the central source and one of them does not closely adjoin the CO emission.

{\bf CH$_3$OH morphology:} Compact methanol emission in the CH$_3$OH (5$_1$--4$_2$) transition is seen in all sources except L1157. In IRAS4B, there is some additional emission $\sim$4$''$ north of the continuum emission peak along the N-S outflow direction.

\subsection{Size measurements}\label{snowlines1}

In order to determine the extent of the emission regions of C$^{18}$O and CH$_3$OH, we performed Gaussian fits in the uv-plane, where we averaged the spectral channels over $\pm$ 3 km s$^{-1}$ around the systemic velocity of each source. In order not to be too much influenced by the irregularities in the emission morphologies, we fixed the centroid of the fits on the source positions. 

The  elliptical fits of C$^{18}$O confirm the elongation along the outflow axis within 15$^\circ$ for all sources except L1448C (see Table \ref{fitsc18o}). Because we want to exclude the perturbing influence of the outflow to probe only the effect of source heating, we also performed fits with the orientation of the ellipse fixed along the outflow directions at the respective P.A. as found in the literature (see Section \ref{thesources} and Table \ref{fitsc18o}). For IRAS4A, we used the small scale value of P.A. = 0$^\circ$ which also excludes the interfering influence of IRAS4A1. For L1448C we used a PA of 160$^{\circ}$, which corresponds to the orientation of the red-shifted component (\citealt{Hirano:2010}). Finally, for comparison, we also performed circular Gaussian fits, even though this source model apparently does not match the observed morphology of the sources at all. 

If we take the values obtained from fits with fixed major axis as being the most reliable fits to trace the "unperturbed" extent of the CO emission, the fitted HWHM along the direction perpendicular to the outflow varies between $\sim$280 and $\sim$560 au in our sources. Based on the luminosity values from Table \ref{sources}, the size of the emission region grows with higher luminosity as expected if CO ice is sublimated mainly by radiative heating by the central source. 

However, we warn that the values derived from elliptical Gaussian fits in the uv-plane need to be taken with some caution.
From the emission maps shown in Fig. \ref{maps}, it is already apparent that the model of an elliptical Gaussian source morphology does not come close to the rather complex emission we see in our high angular resolution data. 
Accordingly, the complex source morphology adds an additional source of uncertainty to the fits beyond the formal error from the fitting procedure itself as reported in Tables \ref{fitsc18o} and \ref{fitsch3oh}.   Plots of the radially averaged uv-data of the C$^{18}$O emission can be found in Appendix \ref{uvplots}.

The compact methanol emission looks better suited to be fit by elliptical Gaussians than the rather complex emission in C$^{18}$O. However, also here the influence of the outflow together with a low signal to noise ratio in IRAS4B and L1448C challenge the quality of the fits. Especially in IRAS4B, the observed methanol emission is strongly affected by the outflow in the form of an extended pedestal, therefore the uv fit does not reliably trace the CH$_3$OH snow line, as can be seen from the large eccentricity of the fit along the outflow (see Table \ref{fitsch3oh}). This problem also affects the trustworthiness of the fit with fixed P.A. The fits for IRAS4A and L1448C show the uv-fitted Gaussian radius  of the CH$_3$OH emission zone at $\sim$140 and $\sim$50 au, respectively. We note, however, that these fits already operate at the limiting scale of our angular resolution and should accordingly be taken with some caution also for that reason. Among the sources, IRAS4A is expected to be best suited for tracing the methanol snow line, due to its high luminosity and the high signal to noise ratio in the methanol transition, even though the P.A. and the eccentricity of the fit hint at an influence of the outflow also in this source. 
Plots of the radially averaged uv-data of the CH$_3$OH emission are shown in Appendix \ref{uvplots}.
It may be interesting to compare our methanol observations with observations of H$_2{}^{18}$O, obtained by \citet{Jorgensen:2010} and \citet{Persson:2012}. Their observations, also performed with the PdBI, have synthesised beam sizes of 0.67$''$ $\times$ 0.5$''$ and 0.86$''$ $\times$ 0.70$''$ for IRAS4B and 4A respectively. The authors find extents of the water emission regions of 0.2$''$ and 0.6$''$ (FWHM values of Gaussian fits)  in 4B and 4A. Accordingly, we see a methanol emission region in IRAS4A that is twice the size of the cited water observation.

Finally, how does the HWHM of Gaussian fits to the observed emission regions relate to the corresponding snow line? 
The snow line marks the radius, at which a particular species starts freezing out onto the dust grains (for a more technical definition, see Sect. \ref{COtemp}). With respect to the abundance profiles, we theoretically expect three different regions: the central region where all molecules are in the gas phase, a transition region right outside of the snow line, where the molecules are partly depleted, and the outer region where all the molecules are found as ices on the surface of the grains. Inside the snow line, the gas-phase abundance is roughly constant, but the density and temperature both increase as power-laws towards the central source. Accordingly the line brightness also strongly increases towards the center, and the emission radius at half- maximum, as measured by Gaussian fits, is expected to be smaller than the true extent of the snow line. How close the observed HWHM of the emission region comes to the radius of the snow line depends on various factors like the width of the transition region, the temperature and the density profiles, and the observing beam. Thus, in order to derive a reliable estimate of the location of the snow line from the uv-fitted HWHM or the full observed intensity profiles, chemical and radiative transfer modelling, and a proper simulation of beam smearing by the PdBI interferometer, is required.

  \begin{table*}[!htb]
\footnotesize
\caption{Results of elliptical and circular Gaussian fits in the uv plane to the C$^{18}$O emission integrated over $\pm$3 km s$^{-1}$,  performed with task \texttt{uv\_fit} in \texttt{MAPPING}. The centroid of all fits was fixed to the position of the dust continuum peak.}            
\label{fitsc18o}      
\centering                          
\begin{tabular}{l l l l l l l l}  
\hline\hline   
\noalign{\smallskip}
\multicolumn{1} {c} {Source}& \multicolumn{1} {c} {P.A.}& \multicolumn {2} {c} {Elliptical fit} &\multicolumn{1}{c}{Circular fit} & \multicolumn{3} {c} {Elliptical fit with fixed P.A. along outflow}  \\
\cline{3-8}
\noalign{\smallskip}
   & &Major $\times$ Minor\tablefoottext{a} (P.A.) & Errors & FWHM  &  Perp. axis   &  Radius\tablefoottext{b} &T$_{\rm dust}$(r)\tablefoottext{c}\\
& ($^{\circ}$)&($''$ $\times$ $''$ ($^\circ$)) & ($''$ $\times$ $''$ ($^\circ$)) & ($''$)  & ($''$)  & (au) & (K)\\
\noalign{\smallskip}
 \hline           
\noalign{\smallskip}
IRAS4A  &0& 6.27 $\times$ 4.62 (-9.7) &0.17 $\times$ 0.08 (2.3)&  5.17 $\pm$ 0.07 & 4.76$\pm$ 0.07 &560 $\pm$ 8 & 28\\
L1448C &160 &6.30 $\times$ 2.00 (34.0) &0.11 $\times$ 0.21 ( 1.0)&  4.31 $\pm$ 0.11 & 4.41$\pm$ 0.14 & 518 $\pm$ 16 &27\\
L1157 &161 &4.85 $\times$ 2.44 ( -32.0) &0.19 $\times$ 0.11 (2.3)&  2.33 $\pm$ 0.06 & 2.61 $\pm$ 0.11 & 424 $\pm$ 18 &26\\
IRAS4B & 0&4.47 $\times$ 2.36 ( -6.1) &0.14 $\times$ 0.07 (1.8)&  3.09 $\pm$ 0.07 & 2.38$\pm$ 0.08 & 280 $\pm$ 9 & 30\\
\noalign{\smallskip}
\hline
\end{tabular}
\tablefoot{
\tablefoottext{a}{Major and minor axes of the elliptical fits correspond to the respective FWHM. }
\tablefoottext{b}{The radius perpendicular to the outflow as derived from the HWHM of the fit with fixed P.A. along the outflow.}
\tablefoottext{c}{The dust temperature at the radius perpendicular to the outflow was determined using the profiles calculated by \citet{Kristensen:2012}.}
}
\end{table*}

  \begin{table*}[!htb]
\footnotesize
\caption{Results of elliptical and circular Gaussian fits in the uv plane to the CH$_{3}$OH emission integrated over $\pm$3 km s$^{-1}$, performed with task \texttt{uv\_fit} in \texttt{MAPPING}. The centroid of all fits was fixed to the position of the dust continuum peak}            
\label{fitsch3oh}      
\centering                          
\begin{tabular}{l l l l l l l l l l}     
\hline\hline
\noalign{\smallskip}
\multicolumn{1} {c} {Source}& \multicolumn {1} {c} {P.A.}& \multicolumn {2} {c} {Elliptical fit} &\multicolumn{1}{c}{Circular fit} & \multicolumn{3} {c} {Elliptical fit with fixed P.A. along outflow}  \\
\cline{3-8}
\noalign{\smallskip}
   & &Major $\times$ Minor\tablefoottext{a} (P.A.) & Errors & FWHM  &  Perp. axis   &  Radius\tablefoottext{b} &T$_{\rm dust}$(r)\tablefoottext{c}\\
& ($^{\circ}$)&($''$ $\times$ $''$ ($^\circ$)) & ($''$ $\times$ $''$ ($^\circ$)) & ($''$)  & ($''$)  & (au) & (K)\\
\noalign{\smallskip}
 \hline           
\noalign{\smallskip}
IRAS4A &0  &1.61 $\times$ 1.15 (19.9) &0.09 $\times$ 0.06 (5.8)&  1.30 $\pm$ 0.05 & 1.20 $\pm$ 0.06 &141 $\pm$ 7  & 70\\
L1448C  &160&0.54 $\times$ 0.34 (-38.3) &0.12 $\times$ 0.15 ( 26.9)&  0.47 $\pm$ 0.08 & 0.37 $\pm$ 0.13 &47 $\pm$ 15  & 110\\
IRAS4B\tablefoottext{d} &0&10.68 $\times$ 1.21 ( 2.4) &0.92 $\times$ 0.11 (0.9)&  1.77 $\pm$ 0.11 & 1.43 $\pm$ 0.13 &168 $\pm$ 15  & 40\\
\noalign{\smallskip}
 \hline           
\end{tabular}
\tablefoot{
\tablefoottext{a}{Major and minor axes of the elliptical fits correspond to the respective FWHM. }
\tablefoottext{b}{The radius perpendicular to the outflow as derived from the HWHM of the fit with fixed P.A. along the outflow.}
\tablefoottext{c}{The dust temperature at the radius perpendicular to the outflow was determined using the profiles calculated by \citet{Kristensen:2012}.}
\tablefoottext{d}{The uv-fits for IRAS4B are strongly affected by outflow contamination and therefore do not reliably trace the central compact methanol emission.}
}
\end{table*}

\section{Analysis}\label{analysis}
\subsection{The model}\label{model}

We aim at reproducing the observed  emission in N$_2$H$^+$, C$^{18}$O, and CH$_3$OH using a chemical model with the envelope density and temperature structure of the respective source, coupled with a simple radiative transfer model.  We will vary the binding energies of CO and N$_2$ ices and  their abundances to fit the observed sizes and intensities in C$^{18}$O and N$_2$H$^+$.

As input for the chemistry model, we use the 1-D density and temperature profiles published by \citet{Kristensen:2012}. They assume that the density profiles follow a power-law with $n \propto r^{-p}$. The inner boundary of the model is set at a radius where the dust temperature is 250 K. Using the DUSTY envelope model, the power-law index $p$ of the density models was constrained for each source using the radial dust emission profiles of 450 and 850 $\mu$m SCUBA observations  tracing the outer envelope. In the production of the radial emission profiles, regions around the source that show irregularities like other sources or bad data were excluded. The spectral energy distributions of the sources as compiled from the literature were used to determine the extent of the envelope and the dust optical depth at 100 $\mu$m. The model parameters for our sources are listed in Table \ref{modelparam}. We assume that the gas and the dust have the same temperature, which is justified at the high densities considered (which are $\sim$10$^3$ cm$^{-3}$ in the outermost envelope and a few 10$^9$ cm$^{-3}$ in the center).

 \begin{table*}[!htb]
\footnotesize
\caption{Model parameters.}            
\label{modelparam}      
\centering                          
\begin{tabular}{l l l l l l l l l l l }     
\hline\hline
\noalign{\smallskip}
Source   & r$_{\rm in}$\tablefoottext{a} & $p$\tablefoottext{a} & $\tau_{100}$\tablefoottext{a} &  db\tablefoottext{b} &X(C$^{18}$O)$_{\rm plateau}$\tablefoottext{c} &X(N$_2$H$^+$)$_{\rm peak}$ &X(CH$_3$OH)$_{\rm plateau}$ &  r$_{\rm snow}$(CO)\tablefoottext{d} & T$_{\rm snow}$(CO)\tablefoottext{d} &  r$_{\rm snow}$(N$_2$)\tablefoottext{e}\\
& (au) &  & & (km s$^{-1}$) & & & & (au) & (K)\\
\noalign{\smallskip}
 \hline           
\noalign{\smallskip}
IRAS4A & 33.47&1.8 & 7.7 & 0.50&  2$\times$10$^{-8}$& 1$\times$10$^{-10}$&4$\times$10$^{-8}$& 770&24&1200\\ 
L1448C & 20.66&1.5 & 3.2 & 0.35 & 2$\times$10$^{-8}$&2$\times$10$^{-10}$&  8$\times$10$^{-8}$ &730&23& 1180\\ 
L1157 & 14.38& 1.6 & 2.5 & 0.80&   6$\times$10$^{-8}$& 6$\times$10$^{-10}$& $<$ 2$\times$10$^{-8}$ &540&24&950\\
IRAS4B & 15.00&1.4 & 4.3 & 0.40 &3$\times$10$^{-8}$& 2$\times$10$^{-10}$&   3$\times$10$^{-7}$ &460&24&760 \\ 
\noalign{\smallskip}
 \hline           
\end{tabular}
\tablefoot{
\tablefoottext{a}{The envelope's inner radius r$_{\rm in}$, the power law index $p$, and the optical depth at 100 $\mu$m $\tau_{100}$ are taken from \citet{Kristensen:2012}.}
\tablefoottext{b}{The Doppler parameter db is obtained by matching the modelled and the observed line widths.}
\tablefoottext{c}{All fractional abundances in this table are given relative to H$_{2}$.}
\tablefoottext{d}{Radius and temperature of the CO snow line as determined from chemical modelling.}
\tablefoottext{2}{Radius of the N$_2$ snow line as determined from chemical modelling. The temperature at these radii is $\sim$19 K for all sources.}
}
\end{table*}

Based on these density and temperature profiles, we used the Astrochem chemistry code (e.g. \citealt{Maret:2015}) to determine the chemistry in the envelopes of our sources. This code computes the abundances of chemical species as a function of time for a stationary, spherically symmetrical source structure, where each shell of the source is at constant density and temperature. The code accounts for gas-phase reactions, cosmic-ray- and photo ionization, and photo-dissociation as well as for gas-grain interactions, namely H$_2$ formation on grains, electron attachment and ion recombination, depletion and desorption processes.  We assume a standard value for the cosmic ray ionization rate of 1.3$\times$10$^{-17}$ s$^{-1}$. The dust grain abundance relative to H$_{2}$ that we use in our model is X$_{\rm d}$ = 2.6 $\times$ 10$^{-12}$, which corresponds to a n$_{\rm H2}$ to n$_{\rm d}$ ratio of 3.79 $\times$ 10$^{11}$. The average grain size is 0.1 $\mu$m, so that the assumed gas-to-dust mass ratio is 100. The average grain size and the grain abundance we use is chosen such that it gives the same total grain surface area per H nucleus as an MRN distribution. 

We use a chemical network containing gas-phase reactions and rates from the Ohio State university (OSU) astrochemistry database in its 2009 version with updated nitrogen reaction rates as published in \citet{Daranlot:2012}. Binding energies and reaction rates for the grain-gas interactions were taken from \citet{Garrod:2006,Oberg:2009a,Oberg:2009b,Bringa:2004}. For the CO and N$_2$ ices in particular, we initially assume multilayer desorption of pure  ices and corresponding binding energies of 855 K and 790 K, respectively (\citealt{Oberg:2005}). We further assume that methanol desorbs at the same temperature as the water ices, as its desorption behaviour, being classified as a "water-like species", is assumed to be dominated by hydrogen-bond interactions with water (\citealt{Collings:2004}). This corresponds to an assumed binding energy for methanol of 5700 K.  We however note that this value constitutes an upper limit, as methanol may also be contained in CO dominated ices, see the discussion in Sect. \ref{ch3oh}. The values of the binding energies determine the size of the respective emission regions, because they determine the dust temperature at which the freeze-out of gas-phase species occurs. Given that the exact value of the binding energies depends on the (unknown) composition of the ices, they are used as free parameters to adjust the modelled emission sizes to the observed ones. We will follow this approach in Section \ref{testcase}. However, we will not attempt to adjust the CH$_3$OH binding energy, since the observed emission sizes in this species appear too uncertain and affected by outflow contamination.

The adopted elemental abundances correspond to the "low metal case" from \citet{Wakelam:2008}, but with a slightly lower oxygen elemental abundance of 3.2 $\times$ 10$^{-4}$ relative to H$_{2}$, following \citet{Hincelin:2011}.  Our chemical initial conditions include six molecules: H$_2$, NH$_3$, CO, N$_2$, CH$_3$OH, and H$_2$O. Except for H$_2$, the molecules were initially put into ices, because the thermal desorption timescale is much shorter than the depletion timescale inside of the snow line (see, e.g., Fig. \ref{T_evap}), and therefore the molecules return quickly to the gas phase in that region. In order to keep the number of free parameters small, we only vary the initial abundances of CO, N$_2$, and CH$_3$OH to match the observed line peak intensities. In all the models, we include, somewhat arbitrarily, 10\% of the elemental nitrogen abundance in NH$_3$ ices.  This value is in line with a total fraction of nitrogen locked up in ices of up to 10\%--20\% as found by \citet{Bottinelli:2010}. The oxygen that is not bound in CO and CH$_3$OH is put in water ice. The water abundance will thus be overestimated, given that we do not include CO$_2$ in the chemical initial conditions in order not to introduce another unknown parameter in our model. This should not have a critical effect on our modelling results of the CO emission, because most of the CO emission stems from a region where water is still bound in ices. By varying the initial water abundance, we checked that also the more compact methanol emission is not affected by the assumed water abundance (see Appendix \ref{chem}). 

We stop our chemical models at a time of $\sim$5 $\times$ 10$^4$ yr, hereafter referred to as chemical age. 
We note that for CO the adaption of the initial CO abundance in order to match the observed intensity is straightforward, as in the central region basically all CO that is initially put into CO ices is finally found in the gas phase with little change over time. For N$_2$H$^+$, we matched the peak intensity found at the ring radius, but here we find a slight time dependence as N$_2$H$^+$ is destroyed over time. For methanol, the final abundance depends rather strongly on the assumed chemical evolution time, because methanol strongly participates in chemical reactions once it is in the gas phase.  Appendix \ref{chem} shows the computed abundance profiles for several chemical ages.

 We stress that the comparison between our model and the observations will only constrain the present-time peak values that are fed into the radiative transfer routine. They cannot constrain however the initial abundances, as the inference to the initial abundances from the observed emission depends on unknown parameters like the chemical age or the specific chemical composition of the gas. We list in Table \ref{modelparam} our best determined parameters, namely the plateau or peak values of the present-time abundance profiles of C$^{18}$O, N$_2$H$^+$ and CH$_3$OH that are obtained as outputs of Astrochem and then used to compute the emission maps that best reproduce the observations.

For the sake of information, we have however also listed all initial abundances that we used in Astrochem to model our four sources in Table \ref{iniabun}.
We stress that the fitted central methanol abundances as listed in Table \ref{modelparam} are to be understood as upper limits, given that the modelled gaussian size (assuming a binding energy of 5700 K) is generally much smaller than the observed uv-fitted size.

The resulting abundance profiles are finally fed into the radiative transfer code RATRAN (\citealt{Hogerheijde:2000}). This code computes the molecular excitation and the resulting molecular line emission based on the Monte Carlo method for an axially symmetric source model. 
In our computations, we assume no radial velocity gradient and a $^{16}$O/$^{18}$O isotopic ratio of 500, consistent with the value of 479$\pm$29 found by \citet{Scott:2006}. The doppler parameter for each source is chosen as to match the width of the C$^{18}$O lines. The corresponding values are listed in Table \ref{modelparam}. 
The molecular data that is needed in the computation of the excitation characteristics is taken from the Leiden Atomic and Molecular Database and include collisional rates from \citet{Yang:2010,Schoier:2005,Daniel:2005,Rabli:2010}. For C$^{18}$O, where different collision rates exist for ortho- and para-H$_2$, we assume that collisions with ortho-H$_2$ dominate in the high temperature regions where the observed radiation originates. Given that we observe optically thin emission in N$_2$H$^+$ (see Sect. \ref{snowlines}), we calculate all hyperfine components separately and then add the intensities of the components.

From the resulting images we then compute uv tables with the GILDAS mapping software, using the same uv coverage as in our observations. Therefore, the effect of missing short spacing information that filters out the large-scale envelope structure in the observations is compensated for in the model and the potentially disturbing influence of large-scale emission from the colder envelope is prevented. Subsequently the same imaging and image deconvolution steps are applied to both the observed and the modelled uv data. In order to bring the model in accordance with the observations, the initial molecular abundances of CO, N$_2$, and CH$_3$OH are used as free parameters (varied in steps of the first decimal place of the abundance) to match the observed peak intensities, while the binding energies were varied (in steps of 100 K) to match the observed emission size in the uv-plane and the observed intensity cuts of the spatially resolved emission.  For the optimization no automated grid approach was used but the matching between model and observation was done by eye. Appendix \ref{chem} illustrates the impact of changes in the CO binding energy as well as the impact of the initial CO abundance on the resulting chemical abundance profiles.

\subsection{The testcase: IRAS4B}\label{testcase}

IRAS4B is the source with the least disturbed anti-correlation pattern in C$^{18}$O and N$_2$H$^+$. Furthermore, it seems to be a single source with relatively simple continuum and C$^{18}$O emission morphologies. Therefore we take this protostar as a test case on which we adjust and explore our modelling approach. 

If we just run the chain of models as described in Section \ref{model} on the envelope structure of IRAS4B, we obtain a C$^{18}$O emission region that is clearly more extended than our observations. A Gaussian fit in the uv-plane of the resulting synthetic observations yields a minor axis (FWHM) in East-West direction of 5.3$''$, more than twice the fitted size of the observation. The predicted intensity cuts perpendicular to the flow are also twice too broad.

The radius where CO starts freezing-out on grains depends on mainly two different factors: the temperature profile of the source model and the CO ice binding energy. A decrease of the temperature profile would reduce the size of the CO emission region but it would also decrease the size of compact methanol emission. We will discuss this scenario, which links back to possible uncertainties in the source luminosity, in Section \ref{Discussion}. Here we only note that the C$^{18}$O morphology in IRAS4B could be reproduced with temperature and density profiles divided by a factor of 1.5 compared to the profiles published by \citet{Kristensen:2012}. However, in the following we adapt the CO binding energy in order to match the observed emission regions.

   \begin{figure*}[!htb]
   \sidecaption
   \includegraphics[width=12.5cm]{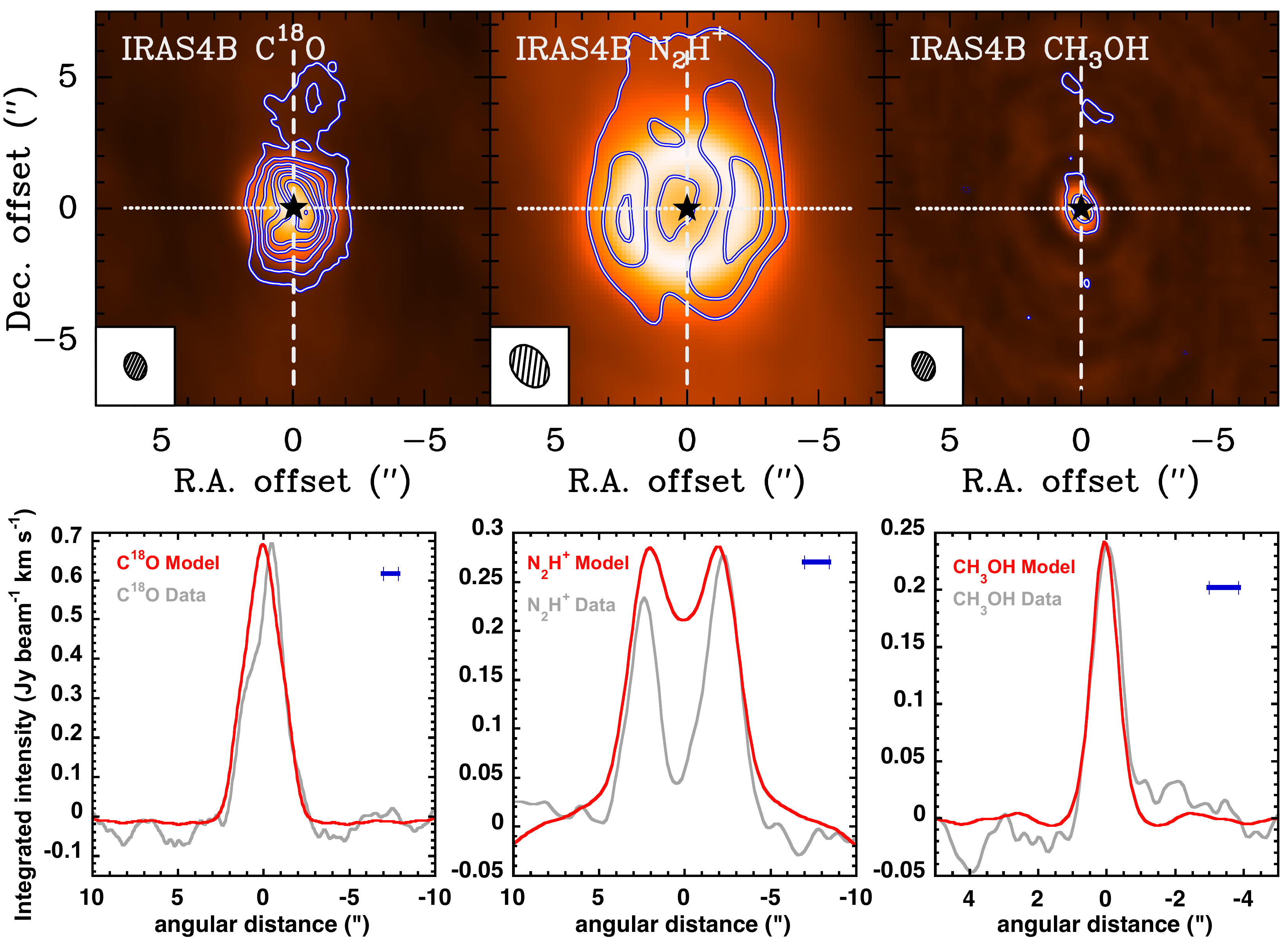}
      \caption{Top row: Comparison of the observations (contours) with the synthetic maps (colour background) produced by the best fit model for IRAS4B with E$_{\rm b}$(CO) = 1200 K and E$_{\rm b}$(N$_2$) = 1000 K. Top left: C$^{18}$O (2--1), centre: N$_2$H$^+$ (1--0), right: CH$_3$OH (5$_1$--4$_2$), integration intervals are the same as in Fig. \ref{maps}, while the contour spacing is in steps of 3$\sigma$, starting at 3$\sigma$. The white dashed lines show the outflow direction, while the dotted lines show the direction of the intensity cuts. The scaling of the maps can be understood in comparison with the cuts in the bottom row. Bottom row: Cuts perpendicular to the outflow direction (P.A. = 0$^\circ$) for C$^{18}$O, N$_2$H$^+$, and CH$_3$OH (left to right). Grey lines show the data, red lines the model. The blue bars in the top right corners of the bottom panels show the HPBW along the cut. 
      }
\label{model_iras4b}
   \end{figure*}

The value of the CO binding energy we initially use in the model is 855 K (\citealt{Oberg:2005}). We can reproduce the observed emission size in our model if we increase that value to 1200 K. 
In Figure \ref{model_iras4b} we also compare the model and the observations in the image domain, where in the upper panel the observations are overlaid as contours on the model as colour background. For better comparison, the lower panel shows intensity cuts from model and observation perpendicular to the outflow direction. The cuts are seen to agree very well.

The match of the N$_2$H$^+$ ring also requires an adaption of the N$_2$ binding energy, because the size of the ring depends on that parameter. While the inner radius of the ring is set by the desorption of CO into the gas phase and corresponding destruction of N$_2$H$^+$, the outer radius of the ring is set by the freeze-out of N$_2$ ices and thus by the N$_2$ binding energy. If we use a value of 1000 K, we obtain a very good match between the synthetic and the observed ring perpendicular to the outflow axis. The only shortcoming is that our model still shows quite some emission towards the center.  Figure \ref{chemistry_dependence} in Appendix \ref{chem} shows that this is not a time-dependent effect. It rather seems like our model misses some mechanism to destroy N$_2$H$^+$ in the innermost envelope. Besides by CO, N$_2$H$^+$ is also destroyed by water, CO$_2$, and free electrons.  However water is released from ices only in the innermost regions, as confirmed by the small H$_2$O sizes observed in two of our sources, and this effect is already included in our model.
At the same time, all the oxygen is bound in CO and H$_2$O in our model. Accordingly we do not account for the effect of N$_2$H$^+$ destruction by CO$_2$ via proton transfer, which occurs in a larger region than the destruction by H$_2$O due to the lower binding energy of CO$_2$ ices compared to water ices. In addition, destruction by free electrons may probably be at work. These could either be created by UV radiation stemming from the central protostellar object and penetrating into the inner envelope inside of the outflow cavity (e.g. \citealt{Visser:2012}), or, if present, by X-rays in the innermost region (e.g. \citealt{Staeuber:2007,Prisinzano:2008}). In the framework of our modelling approach, we are however not able to disentangle these possible effects.

In the case of methanol, the comparison between model and observation in the uv-plane is hindered by the outflow contamination of the observations and by the resulting problems to apply a Gaussian model. Therefore we refrained from attempting to fit the observed size with a modified methanol binding energy and kept the same value of 5700 K as for water ices.
The model fit yields a FWHM of 0.2$''$ for IRAS4B. This is much smaller than the average FWHM of complex organic molecule emission in this source of 0.5$\pm$0.15$''$, as we observe it in our WideX spectra (Belloche et al. in prep.). In the image domain, model and observation have the same intensity profile corresponding to the beam (see Fig. \ref{model_iras4b}), but this merely illustrates that the modelled and actual emission sizes are smaller than the beam size.

\begin{figure*}[!htb]
   \sidecaption
   \includegraphics[width=12.5cm]{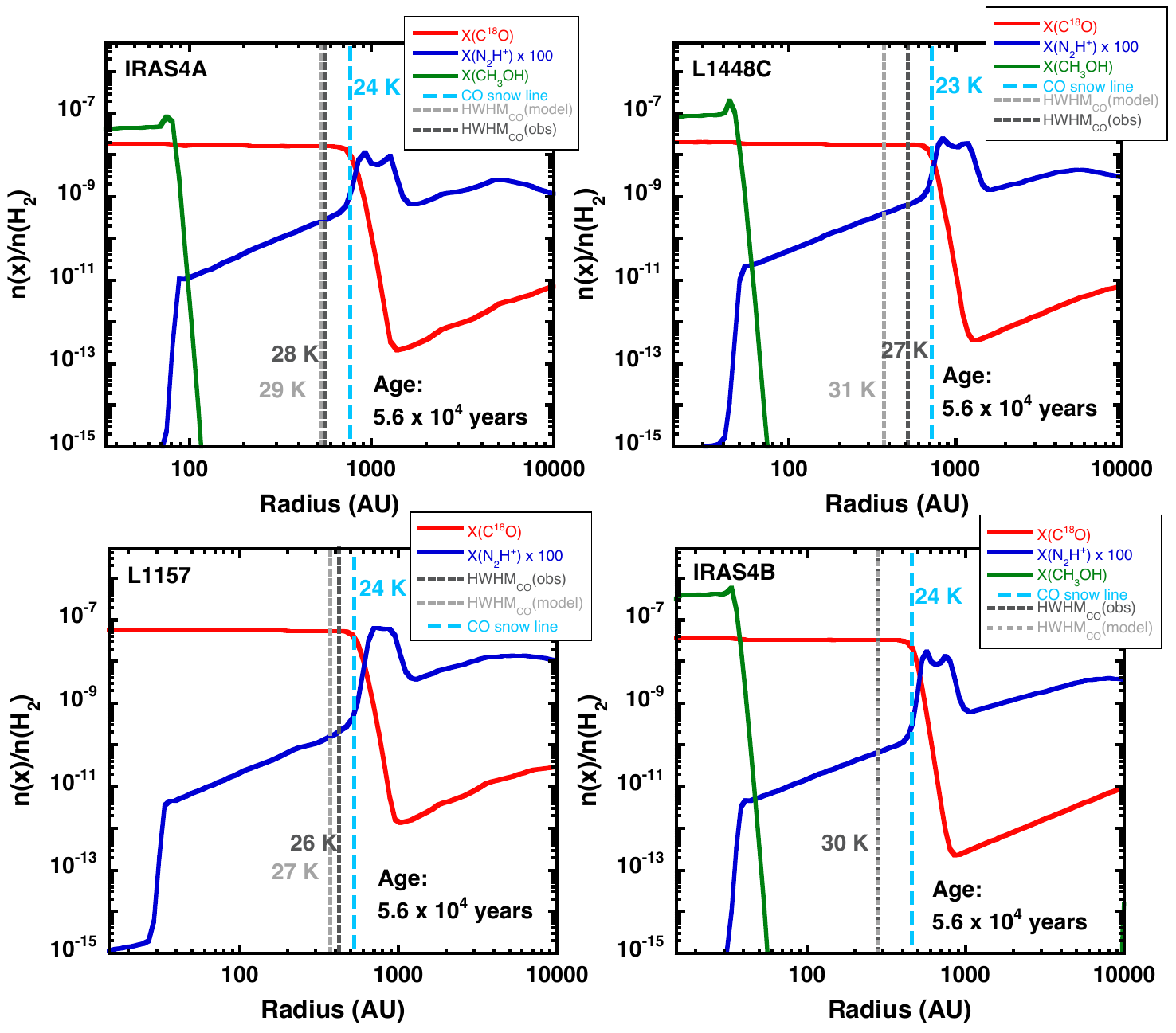}
      \caption{ Fractional chemical abundances, relative to H$_{2}$, of C$^{18}$O (red lines), N$_2$H$^+$ (blue lines), and CH$_3$OH (green lines), computed by Astrochem for the envelope structures of IRAS4A (top left), L1448C (top right), L1157 (bottom left), and IRAS4B (bottom right). The light grey dotted lines show the HWHM of the modelled emission, the dark grey dotted lines the HWHM of the observed emission in C$^{18}$O (see Table \ref{fitsc18o}), while the light blue dashed line shows the location of the CO snow line in the model (see Table \ref{modelparam}). The figure also indicates the dust temperature values at the respective radii. In IRAS4B, the observed and modelled HWHM are the same.
      }
\label{chemistry_iras4b}
   \end{figure*}

The underlying chemical abundances of the model shown in Fig. \ref{model_iras4b} are presented in the lower right panel of Figure \ref{chemistry_iras4b}. As outlined in Sect. \ref{snowlines1}, the abundance profiles of C$^{18}$O and CH$_3$OH show an inner plateau-region, a transition region, and an outer region of depletion. The transition from complete depletion to the maximum gas-phase abundance happens within a few hundred au for CO and about ten au for methanol. The initial abundances of CO, N$_2$, and CH$_3$OH are adjusted as to match the observed  peak intensities (see Table \ref{modelparam} and Table \ref{iniabun}). This procedure yields an inner abundance of C$^{18}$O of about 3$\times$10$^{-8}$, which corresponds to an inner abundance of C$^{16}$O of 1.7$\times$10$^{-5}$ relative to H$_{2}$. The fitted central methanol abundances as listed in Table \ref{modelparam} are to be understood as an upper limit, since the modelled CH$_3$OH Gaussian size is much smaller than observed. If the real extent (and beam filling-factor) of the methanol emission is larger than in our model, the required abundance to match the observed flux would be lower.

In this model, the C$^{18}$O and methanol snow lines, where the respective desorption and depletion rates are equal and where the plateau-abundances start dropping, are located at radii of about 460 au and 35 au, respectively. The N$_2$ snow line is found at a radius of 760 au. In Figure \ref{chemistry_iras4b}, the CO snow line radius is indicated as light blue, dashed line together with the dust temperature at this location, which is $\sim$24 K. 
The HWHM radii of the observed and modeled C$^{18}$O emission, as inferred from Gaussian uv-fits, are also shown as dark grey and light grey vertical dotted lines, respectively. They are identical in IRAS4B, hence superposed in this graph. The corresponding dust temperatures are also indicated. It may be seen that the HWHM is substantially smaller than the true CO snow line radius, by a factor 1.6. Hence, the uv-fitted HWHM of C$^{18}$O only gives an upper limit to the freeze-out CO dust temperature of 30 K instead of 24 K in the present case. The angular radius of the CO snow line in IRAS4B (about 2$''$) corresponds more closely to the radius at zero intensity of the C$^{18}$O intensity cut. However, this coincidence may not hold for all sources or species. For example, the modeled angular radius of the N$_2$ snow line in IRAS4B (about 3$''$) falls closer to the half-maximum of the N$_2$H$^+$ emission cut than to its radius at zero intensity (5$''$). This is because the intensity profile will depend not only on the snow line location but also on the width of the transition region and on the residual abundance outside of it. 

\subsection{The other sources}\label{othersources}

\begin{figure*}[!htb]
   \sidecaption
   \includegraphics[width=12.5cm]{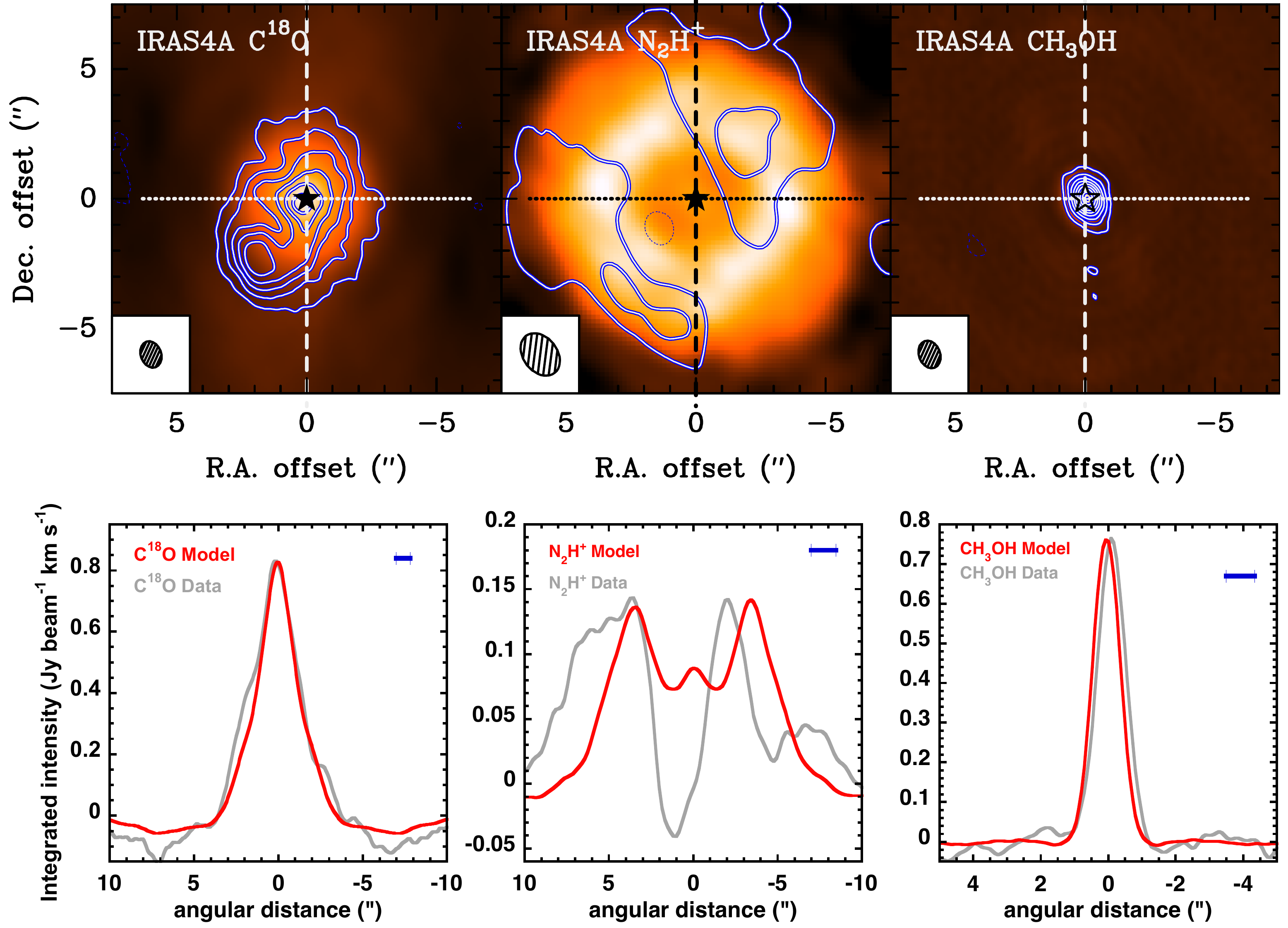}
      \caption{Same as Fig. \ref{model_iras4b}, but for IRAS4A. The cuts are performed at a P.A. of 90$^{\circ}$ perpendicular to the outflow direction (P.A. = 0$^\circ$).}
\label{model_iras4a}
   \end{figure*}

\begin{figure*}[!htb]
   \sidecaption
   \includegraphics[width=12.5cm]{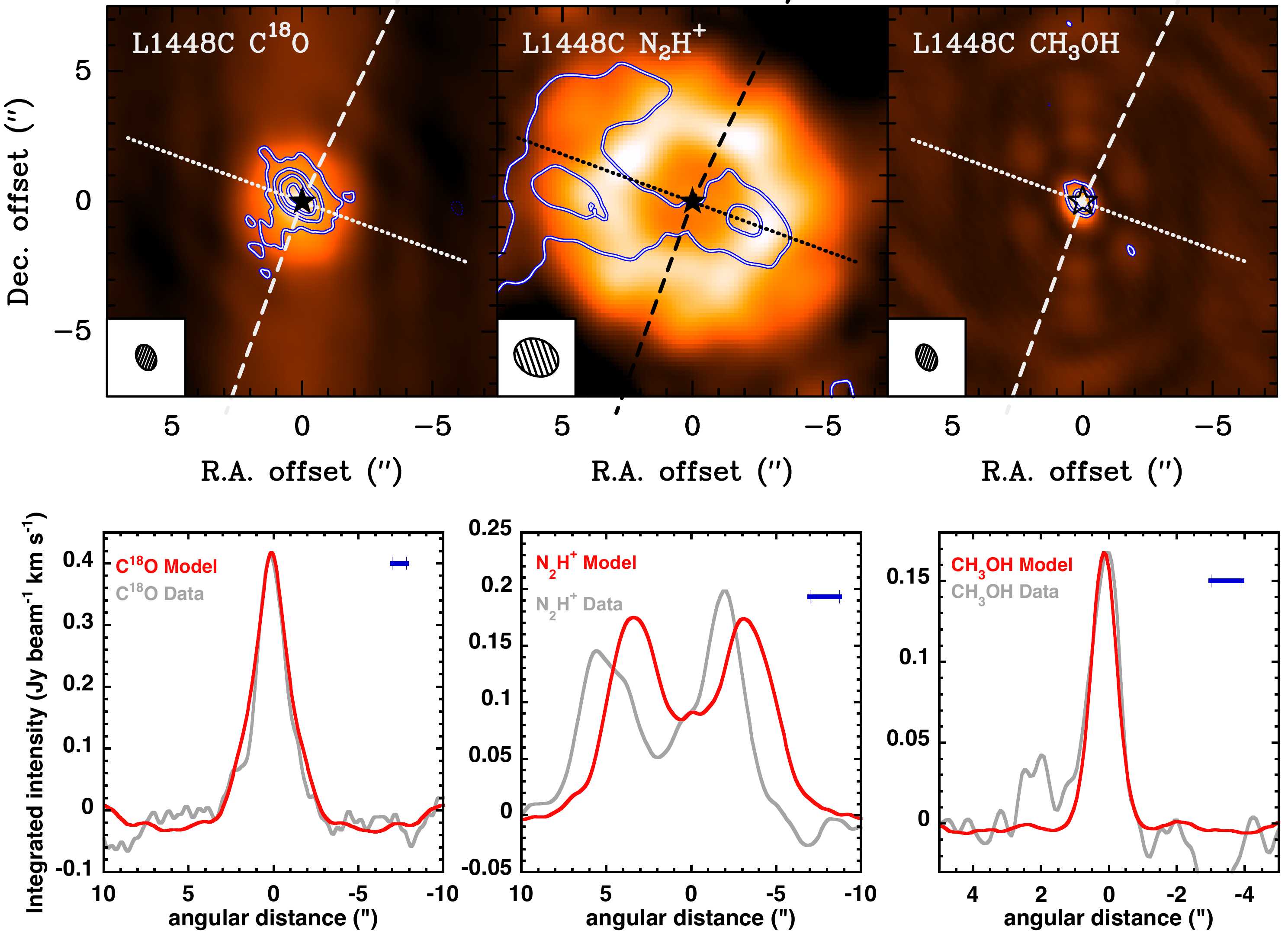}
      \caption{Same as Fig. \ref{model_iras4b}, but for L1448C. The cuts are performed at a P.A. of 70$^{\circ}$ perpendicular to the outflow direction (P.A. = 160$^\circ$).}
      \label{model_l1448c}
   \end{figure*}

\begin{figure}[!htb]
   \centering
      \includegraphics[width=0.48\textwidth]{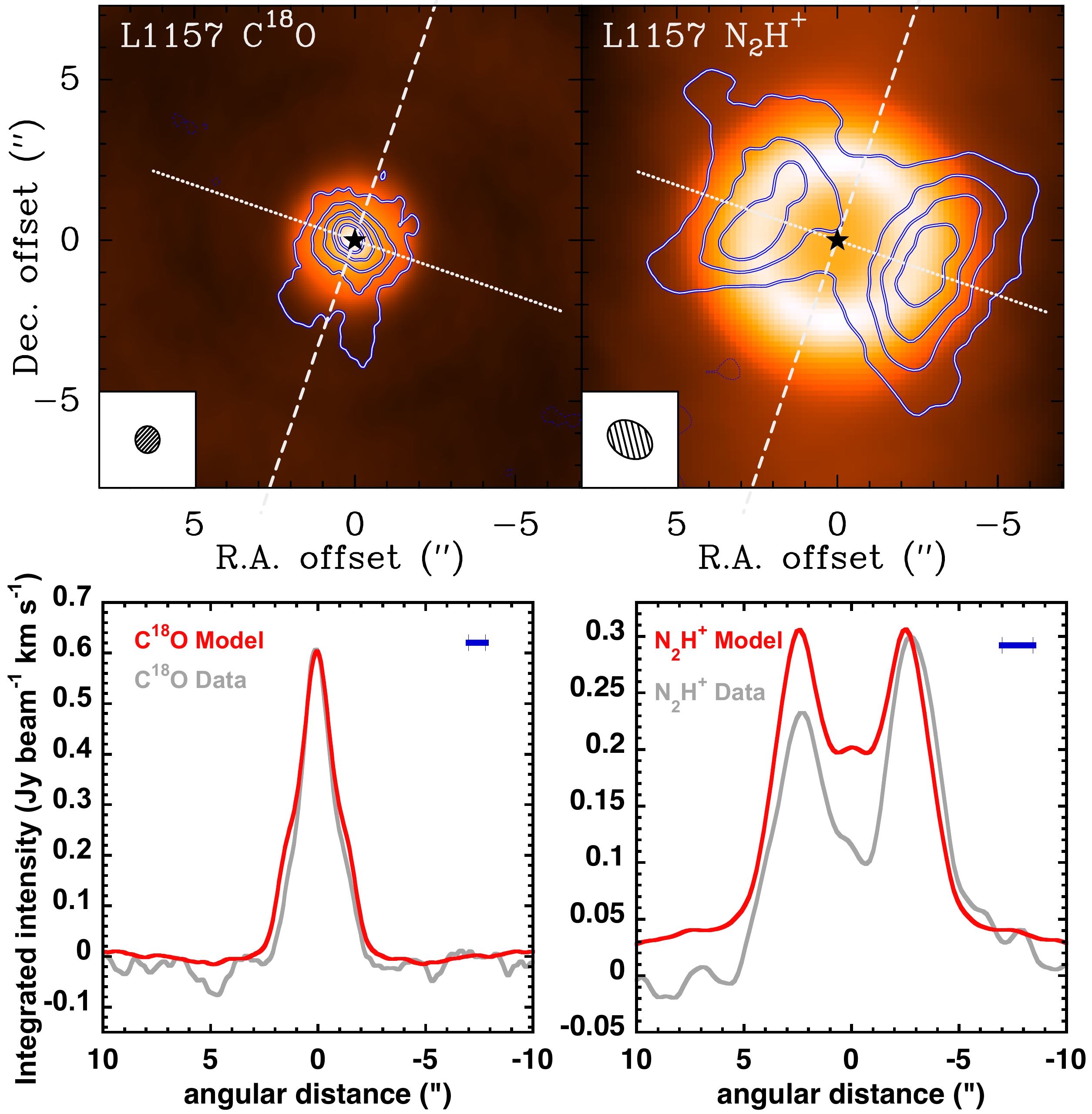}
       \caption{Same as Fig. \ref{model_iras4b}, but without methanol, for L1157. The cuts are performed at a P.A. of 71$^{\circ}$, perpendicular to the outflow direction (P.A. = 161$^\circ$).}
\label{model_l1157}
   \end{figure}

For the other three sources, we used the same values of the binding energies for CO and N$_2$ as were found to reproduce the emission of IRAS4B. The chemical abundance profiles are shown in Fig. \ref{chemistry_iras4b}, and the modelled emission for the three sources is presented in Figs. \ref{model_iras4a}, \ref{model_l1448c}, and \ref{model_l1157}. In all sources, the emission intensity profile of C$^{18}$O is well reproduced, as well as the overall peak location and thickness of the N$_2$H$^+$ ring. 

In IRAS4A the size of the C$^{18}$O emission obtained from Gaussian fits in the uv-plane is a bit smaller for the model than the observed size (4.5$''$ versus 4.8$\pm$0.07$''$). The match of the intensity cuts in the image plane is however satisfactory with an inner C$^{18}$O abundance of $\sim$2$\times$10$^{-8}$ relative to H$_{2}$, which corresponds to an inner abundance of C$^{16}$O of  9$\times$10$^{-6}$. The emission of N$_2$H$^+$ appears strongly disrupted by the influence of the precessing outflow, therefore it is difficult to evaluate the quality of the fit based on one single cut. In the cut displayed in Fig \ref{model_iras4a}, the synthetic observations do not fit the observations very well, but the edges of the emission ring as well as the part of the ring at positive offsets satisfactorily correspond to the modelled ring. We note that in this source, as well as in L1448C, the deconvolution of the modelled observations has to account for the fact that the first sidelobes of the synthesised beam have about the same size as the modelled ring emission. Therefore, in these two sources we have applied a ring-like support in the deconvolution of the models. This support influences the value of the image peak intensity. Accordingly, in these two sources the N$_2$H$^+$ abundance is less well constrained. Again, the methanol emission is reproduced in the image domain at a size corresponding to the beam size. In the uv domain, however, the modelled emission FWHM size is much smaller than the observation (0.44$''$ versus 1.2$\pm$0.06$''$). 

   In L1448C, the intensity cut of the C$^{18}$O emission regions of the model and the observations agree satisfactorily in the image domain perpendicular to the outflow (Fig. \ref{model_l1448c}). The model in N$_2$H$^+$ underscores the rather strange observed morphology, where the emission peaks are not equidistant to the central source: model and observations seem to be shifted with respect to one another, even though the width of the emission region as well as the spatial distance between the peaks seem rather well reproduced. The inner C$^{18}$O abundance in L1448C is found being $\sim$2$\times$10$^{-8}$ relative to H$_{2}$, which corresponds to an inner abundance of C$^{16}$O of 10$^{-5}$. The modelled emission FWHM size for methanol is 0.33$''$, which lies within the error bars of the observed FWHM perpendicular to the outflow of 0.37$\pm$0.13$''$.

For L1157 the match between model and observations in C$^{18}$O and N$_2$H$^+$ is also satisfactory. The distance between the N$_2$H$^+$ maxima is a bit smaller in the model than for the observation. On the other hand, the modelled C$^{18}$O emission is a bit more extended than the observed emission. L1157 has the highest inner C$^{18}$O abundance of the four sources with a value of  6$\times$10$^{-8}$ relative to H$_2$, which corresponds to an inner abundance of C$^{16}$O of  3$\times$10$^{-5}$.

Figure \ref{chemistry_iras4b} shows the chemical abundance profiles for all sources, which are similar in all cases. As already described in Section \ref{testcase} in the case of IRAS4B, the HWHM derived from the Gaussian fit of the C$^{18}$O observation in the uv plane (shown as dark grey, dotted lines) is significantly smaller than the snow line radius (light blue dashed line) in all sources. Notably, this also holds for the relation between the radius of the snow line and the HWHM of the modelled emission, which is a factor of 1.5--2 smaller than the snow line radius.  

We finally note that we also applied our model on IRAM 04191, in order to check whether the locations of the outer and inner radii of the N$_2$H$^+$ ring corresponds to the radii of CO and N$_2$ freeze-out. If we use the temperature and density profiles presented in \citet{Belloche:2002}, according to our model the peak of the N$_2$H$^+$ ring should be located at a radius of $\sim$100 au, while the observations show the ring at a radius of $\sim$1400 au, assuming a distance to IRAM 04191 of 140 pc. Using the same CO and N$_2$ binding energies as for the other sources, the observed ring can only be reproduced with a temperature profile that is three times higher than derived in \citet{Belloche:2002}.  The observed, weak, compact C$^{18}$O emission, however, does not fill the large N$_2$H$^+$ ring, as would be expected if the ring is due to a chemical anti-correlation.
Accordingly, these observations call for further investigation and may suggests that, provided that the temperature and density profiles are correct, the N$_2$H$^+$ ring in IRAM 04191 hints at a different situation than in the other four sources (see also \citealt{Belloche:2004}). It might, e.g., be related to a past accretion burst. We will discuss this source in a forthcoming paper.

 \section{Discussion}\label{Discussion}
 
 As was shown in Section \ref{analysis}, our model is able to reproduce the observed emission well in all four sources if we increase the CO binding energy to a value of 1200 K and the binding energy for N$_2$ ices to 1000 K. The observational fact 
that the CO emission is less extended than what is expected with the initially assumed binding energy for pure CO ices contradicts a scenario, where the chemistry traces a past accretion burst (e.g. \citealt{Visser:2015}). Furthermore, we note that in all cases we need very low CO abundances compared to the general dense ISM (where typical values of the CO abundance in molecular clouds are $\sim$0.6-2 $\times$10$^{-4}$, e.g., \citealt{Ripple:2013}) to reproduce the peak intensities. In this section, we will first address general uncertainties in our modelling approach, before we discuss in more detail our findings on the CO binding energy, the CO freeze-out temperature, the N$_2$ and the methanol binding energies, and the CO abundances. Finally we will briefly compare our findings with a recent study on the CO snow lines in three of our objects by \citet{Jorgensen:2015}.

 \subsection{Uncertainties in our modelling}\label{uncertainties}
 
Our modelling relies on a number of assumptions. One major ingredient are the adopted temperature and density profiles that were derived by \citet{Kristensen:2012}. In order to stay consistent with their modelling, we have used the same values for source distances, luminosities and envelope masses that they assume in their work.

Uncertainties in the source luminosities translate into an uncertainty of the temperature profiles. With a different temperature profile, the same observed position of the snow line would result in a different binding energy. In particular, Fig. 4 of \citet{Jorgensen:2015} shows that we could reproduce our observations with a lower value of the CO binding energy of 855 K if the luminosity was a factor of 4 smaller than the values we assume in our analysis. The uncertainties in the luminosities reported in Sect. 3.1 for L1448C and L1157 are clearly smaller than this factor of 4. Accordingly, the modelling of our observations always requires for these two sources a higher binding energy of CO than the value of multilayer desorption for pure CO ices within the range of uncertainties in source luminosity.  However, as mentioned in Sect. 3.1, the internal luminosities of IRAS4A and IRAS4B derived from {\it Herschel} 70 $\mu$m data are 3.4 L$_\sun$ and 1.5 L$_\sun$, respectively, which are a factor of $\sim$3 smaller than the bolometric values we use in Sect. 4, derived by \citet{Karska:2013}. If the {\it Herschel} internal luminosities were used, we would need to assume a lower CO binding energy  in IRAS4A and 4B  than in L1448C and L1157, implying a different ice composition. We note that adopting bolometric luminosities for all sources, as we have done, gives equal CO ice binding energies for all of them.

The temperature profiles are also subject to other uncertainties beyond variation of the source luminosity only. In their paper, \citet{Kristensen:2012} do not provide estimates for these uncertainties, but in a detailed description of their modelling\footnote{\url{https://github.com/egstrom/Dusty}} they exemplarily show for a Class I source that the temperature profiles as a function of radius are very robust with respect to changes in the power-law density slopes, and variations in the sub-mm and mm fluxes at radii larger than $\sim$100 au. The uncertainty in the temperature profiles, however, grows strongly for smaller radii. Accordingly, while the determination of the CO freeze-out temperature that is reached at radii of a few hundred au should be rather robust (given the luminosity values are basically correct) the uncertainty is larger for the determination of the inner CO abundance, which is affected by the innermost temperature structure and corresponding excitation characteristics.

It is important to note that a different temperature profile, contrary to a change of the CO binding energy, will also affect the size of the methanol emission region. In IRAS4A, the source with the most reliable uv fit of the methanol emission region, we observe a FWHM of the emission region of 1.2$\pm$0.06$''$. With the temperature profile from \citet{Kristensen:2012}, this FWHM value would hint at a freeze-out temperature for methanol of less than 70 K, which already seems unreasonably low given laboratory measurements (e.g. \citealt{Collings:2004} find a binding energy for methanol on water ice of 5530 K, which corresponds to dust temperatures of $\sim$120 K). A reduced temperature profile would decrease the dust temperature at the observed HWHM even further. 
However, we have to stress again that the uncertainty of the temperature profile increases strongly in the central region of the envelope where methanol is present in the gas phase, such that the use of methanol for probing the adopted temperature profile might be rather questionable, all the more as the influence of the outflow on the methanol emission might dominate the effect of source heating (see Sect. \ref{ch3oh}).

There is another uncertainty regarding the distance of L1157, where we use a value of 325 pc, which is by a factor of 1.3 larger than the value of 250 pc that is used, e.g., by \citet{Codella:2015} and Podio et al. in prep. However, the variation of the temperature in the envelope as a function of the angular size should be independent of the assumed distance. This can be seen from equation (2) of \citet{Motte:2001}, which says that the temperature in the envelope as a function of its radius is proportional to (L$_{\star}$ / r$^2$ )$^{0.2}$. Because the luminosity L is given as the product of the observed stellar flux times the distance squared, and the radius r is proportional to the distance times the angular size, the distance dependence cancels out in the cited equation.

Another simplification in our model concerns the initial abundances adopted in Astrochem. While we use the "low metal case" from \citet{Wakelam:2008} for the metals, we put all the carbon in CO and methanol ices and all the nitrogen in N$_2$ and NH$_3$ ices. The CO and N$_2$ ice abundances are then treated as free parameters. This is  a situation that does not correspond to the chemistry we find in dark clouds, where e.g. we also find nitrogen in the form of atomic N (e.g. \citealt{Maret:2006,Daranlot:2012}). Accordingly, as already described in Section \ref{model}, while we give the initial abundances we use in Astrochem in Table \ref{iniabun} for the sake of completeness, the more significant information lies in the abundance profiles that are used in Ratran and that reproduce our observations (see Table \ref{modelparam}).

The simplifications in the initial abundances together with the fact that we use a stationary model also suggest to take the chemical age of our models with some caution. Theoretically, an upper boundary of the evolution time is given by the assumed lifetime of the protostar. As another limiting factor we find that methanol, once  released to the gas phase, is converted by gas phase reactions (the dominant reaction being \texttt{ H$_3$O$^+$ + CH$_3$OH $\rightarrow$ CH$_5$O$^+$ + H$_2$O)}, and its abundance is severely reduced in the center of the envelope typically after about $\sim$1$\times$10$^5$ yr. However, the fact that the envelope is collapsing with free-fall times at the radius of the methanol snow line of a few hundred years, will prevent the effect of methanol destruction from becoming relevant as the envelope material is constantly replenished. Our stationary model does not account for the ongoing collapse, the computed chemistry is however not affected by the collapse, as it is dominated by the effect of ice sublimation that takes place on very short timescales inside of the snow line (see Fig. \ref{T_evap}).  An order of magnitude estimate  of the effect of infall motions on the location of the snow line  is performed in Section \ref{COtemp}. Furthermore, we note that the computed abundance profiles of CO and N$_2$H$^+$ barely change up to a chemical age of $\sim$1$\times$10$^5$ yr. Therefore we do not consider the exact chemical age of the models as being a crucial parameter for our modelling if it is below that value. 

Another caveat of our model is the assumption of spherical geometry, while the observed sources deviate from this assumption due to strong outflow activity and binarity, as we have discussed in Sect. \ref{snowlines}. We account for this asymmetry by comparing observations and models in the direction perpendicular to the outflow direction. 
However, the models do not include the influence of the outflow cavities and related shock and PDR-like features on the excitation of the observed molecules. We note that including this extra heating would tend to broaden the predicted emission sizes. It would thus require an even larger ice binding energy to keep the emission size as small as observed.
 
 \subsection{The CO binding energy}\label{EbCO}
 
Our models can reproduce the observations if we use a CO binding energy of 1200 K. This value is higher than the binding energies of pure CO ices. Laboratory studies differentiate on whether CO ices are located on top of water ices (monolayer desorption) or whether they are bound to a surface of CO ice (multilayer desorption). This differentiation can be understood in the framework of a simple onion model (e.g. \citealt{Collings:2003}), where an inner hydrogenated, water dominated layer is surrounded by an outer dehydrogenated layer dominated by species like CO, CO$_2$, N$_2$ and O$_2$. Accordingly, depending on the thickness of the outer layer, the CO ice is either separated from or directly bound to the underlying water ice, which results in different binding energies and desorption at different temperatures.

However, a number of studies (e.g. \citealt{Collings:2003,Collings:2004,Fayolle:2011,Martin:2014}) have pointed to the fact that the physics and chemistry of the desorption of CO ice layers on grains is more complicated than a simple onion model would suggest. Actually, several desorption features seen at different temperatures show that not only multi- and monolayer desorption play a role for the release of CO from ices to the gas phase.  \citet{Collings:2003} argue that CO can  get trapped in water ice. This is because of a phase change in the water ice component with increasing temperature, which changes the water ice's structure from highly porous amorphous to a less porous amorphous phase. If CO diffuses into the porous structure of water ice at lower temperatures (see, e.g., \citealt{Lauck:2015}), this phase change may then prevent the CO from escaping through surface pores when the ice is heated. Only when the water crystalizes at even higher temperatures, the trapped CO gets desorbed in a process called molecular vulcano. A fraction of the CO that remains in the crystaline water layer is co-desorbed at even higher temperatures together with the water. This behaviour has been confirmed in experiments with precometary ice analogs (\citealt{Martin:2014}). 

Our chemical model does not properly account for the complex layering of ices, but assumes that desorption takes place at a single temperature as described by equations (\ref{eq1}) and (\ref{eq2}). However, in locating the snow lines, being the radii where CO starts freezing out, we are mostly interested in the dominant desorption processes that occur at the lowest temperatures. In the simplistic onion model, these would be multilayer (CO on CO) or monolayer (CO on H$_2$O) desorption, depending on the thickness of the CO containing ice layer.

If we use an average grain size of 0.1 $\mu$m and assume 3 $\times$ 10$^{15}$ cm$^{-2}$ binding sites on the grains' surfaces, the low gas-phase CO abundances in our sources (see Table \ref{modelparam}) correspond to $\sim$2 CO ice layers in IRAS4B and L1157 and $\sim$1 layer in IRAS4A and L1448C. These numbers depend of course strongly on the assumed average grain radius and on the actual layer-structure of the grain ices. However, based on the above values, our models seem consistent with monolayer desorption of CO from water surfaces.

The measurement of the surface binding energies of molecular ices is usually performed using temperature programmed desorption, where ices are grown on a substrate in an ultra-high vacuum chamber. The substrate is then heated and the liberated gases are measured using a mass spectrometer. Using this technique, the binding energy of CO on CO was measured as lying between 850 and 1000 K (e.g., \citealt{Sandford:1988}: 960 K;  \citealt{Bisshop:2006}: 855 K; \citealt{Cleeves:2014}: 855 K; \citealt{Martin:2014}: 890 K), while the binding energy of CO on an amourphous water surface is higher at values between 1200 K and 1700 K (\citealt{Collings:2004}: 1150 K; \citealt{Noble:2012}, coverage of 0.1: 1307 K, coverage of 0.2: 1247 K; \citealt{Cleeves:2014}: 1320 K). 

Indeed, the measured binding energies of CO ices on a water ice surface agree with what we require in our models. Alternatively, if the grain ices in our sources do not exhibit a clean layer structure, the CO binding energy could also be increased with respect to pure CO ices, if the CO ice is mixed with other polar ice components, like e.g. CO$_2$ and organic molecules. \citet{Cleeves:2014} find a binding energy of 1100 K for CO on a CO$_2$ surface, which is also close to the value we find. Indeed, the infrared CO and CO$_2$ ice band shapes and correlations of CO/CO$_2$/H$_2$O ice columns across various lines of sights indicate that about 80\% of CO ice in molecular clouds and towards young stellar objects is in a non-polar CO$_2$:CO mixture (cf. \citealt{Chiar:1994,Pontoppidan:2003,Pontoppidan:2006,Pontoppidan:2008,Whittet:2009}). Also, \citet{Penteado:2015} argue for a gradient of ice mantle composition in grains around young stellar objects where CO is mixed with CH$_3$OH instead of water.

We conclude that while our models seem consistent with monolayer desorption of CO from water ice surfaces,  the structure of the grain ice layers may deviate from a clean layered structure where a pure CO layer sits on top of a layer of water ice. In any case, the value we derive for the binding energy of CO ices agrees well with laboratory measurements. 

\subsection{The CO freeze-out temperature}\label{COtemp}

Observational studies often cite the freeze-out temperature instead of the binding energy. The relationship between the two values in our model, which is based on first-order kinetics, can be understood based on the underlying microphysics of the snow line.
The snow line is located at a radius where the microphysical depletion rate equals the thermal desorption (\citealt{Bergin:1995,Hasegawa:1992}), which translates to
 \begin{equation}\label{eq1}
 S\pi \langle r_d\rangle^2 \varv_{\rm th} n_{\rm d} = \nu_0 \exp{\bigg(-\frac{E_{\rm b}}{T_{\rm d}} \bigg) }
 \end{equation}
 with
 \begin{equation}\label{eq2}
 \varv_{\rm th} = \bigg( \frac{8 k_{\rm B} T_{\rm g}}{\pi m} \bigg)^{1/2} \quad \textrm{and}  \quad  \nu_0 = \bigg( \frac{2N_{\rm S} k_{\rm B} E_{\rm b}}{\pi^2 m} \bigg)^{1/2}  \qquad \textrm{,}
 \end{equation}
 where $ \varv_{\rm th}$ is the thermal velocity, $S$ the sticking probability, which is taken to be 1, $\langle r_{\rm d} \rangle$ the average grain radius, $n_{\rm d}$ the total grain density,  $T_{\rm d}$ the dust temperature, $T_{\rm g}$ the gas temperature, $m$ the mass of the accreting species, and where $\nu_0$ is the characteristic vibrational frequency of the desorbing species with the binding energy $E_{\rm b}$ in Kelvin, and $  N_{\rm S} $ being the number of sites per unit surface. 
 
\begin{figure}[!htb]
   \centering
   \includegraphics[width=0.4\textwidth]{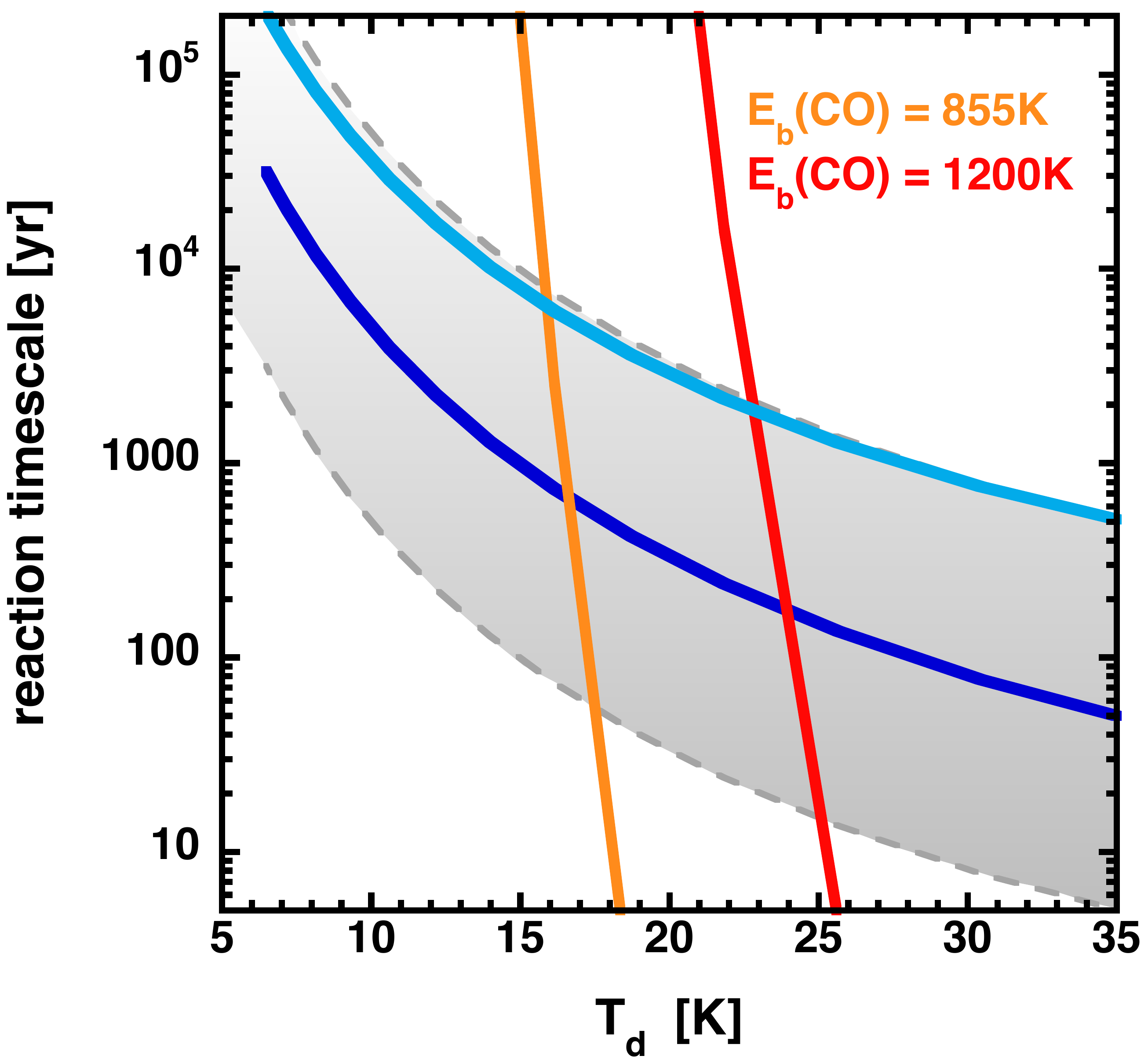}
      \caption{CO ice depletion timescale (dark blue line) and thermal desorption timescales for binding energies of 855 K (orange line) and 1200 K (red line) as function of the dust temperature for the density profiles of IRAS4B.  The light blue line shows the free-fall time for an enclosed mass of 0.5 M$_\sun$. The points of intersection at $\sim$17 K and $\sim$24 K mark the locations of the CO snow line for the two different values of the binding energy. The grey area shows the range of depletion timescales if the dust density is varied by a factor of ten. For a binding energy of 855 K, the variation of the density translates into a variation of the snow line location at $\sim$16 K or $\sim$17 K, while for a binding energy of 1200 K, the modified snow line locations would be at $\sim$23 K or $\sim$25 K.}
      \label{T_evap}
   \end{figure}

 From equation (\ref{eq1}) it is apparent that the exponential in the binding energy is the dominating factor in determining the freeze-out temperature, as can also be seen in Fig \ref{T_evap}, which shows a plot of the timescales in the equation applied to the density and temperature profiles of IRAS4B. Even if the value of the thermal depletion rate is changed, e.g. by varying the total grain density by a factor of 10, the location of the snow line in terms of the dust temperature $T_{\rm d}$ changes by less than 5\%. Figure \ref{T_evap} shows that our best fitting binding energy of 1200 K corresponds to a freeze-out temperature of $\sim$24 K, which is in line with previous observations of protostellar systems. 
 
 Observations indicate that the freeze-out temperature of CO ices depends on the studied environments. Observations of more evolved proto-planetary disks find CO snow lines at lower temperatures of about 17 to 19 K, consistent with a binding energy of 855 K (e.g. \citealt{Qi:2011}: 19 K, \citealt{Mathews:2013}: 19 K, \citealt{Qi:2013}: 17 K), observations of low-mass protostars yield higher values between 20 and 40 K, in line with our findings (e.g. \citealt{Jorgensen:2004}: 40 K, \citealt{Yildiz:2012}: 25 K, \citealt{Jorgensen:2013}: 30 K, \citealt{Jorgensen:2015}: 30 K). \citet{Jorgensen:2015} give a possible explanation of this difference  in terms of grain ice growth. Because grains in disks around T Tauri stars may have been subject to heating and sublimation, the grains may have a clearer onion-like shell structure with a more purified CO ice layer than in protostellar environments, where different ices are probably formed simultaneously at low temperatures.

Based on the desorption reaction timescale displayed in Figure \ref{T_evap}, we can now also obtain an estimate on whether the observed radius of the snow line could also be caused by infalling material with a binding energy of 855 K. This would be the case if the infall timescales are smaller than the desorption timescales. For a binding energy of 855 K, the desorption timescale at a radius of 1200 au where the dust temperature is 16 K is about 600 yr in IRAS4B. At the radius of the observed snow line at 460 au, this timescale has already decreased to about one hour. If we consider the free-fall speed from infinity onto a central mass, $\varv=\sqrt{2GM_*/R}$, we can estimate the free-fall time at radius $R$ as
\begin{equation}
t_{\rm ff}=\sqrt{\frac{R^3}{2GM_*}} \simeq 5035 \, \Bigg( \frac{R}{1000\, {\rm au}} \Bigg)^{3/2} \Bigg( \frac{M_*}{0.5\, M_\sun} \Bigg)^{-1/2} \, {\rm yr} 
\end{equation}
with the gravitational constant $G$ and the mass $M_*$ enclosed inside of radius $R$. Based on this formula\footnote{This expression for the free-fall time is smaller by a factor of 1.6  than the free-fall time $t_{\rm ff,Hunter}=\sqrt{3\pi/(32G\rho)}$ derived for gravitational collapse by \citet{Hunter:1962}, where a constant envelope density $\rho$ is assumed. However, the steep power-law density p = 1.4 in the envelope and the presence of COM emission towards the central source indicate that the early collapse phase described by \citet{Hunter:1962} is over and that a sizeable point mass has formed in the centre that now dominates the gravitational field on the small scales considered.} 
 with an estimated total enclosed mass of 0.5 M$_\sun$, the free-fall time at the radius  of 1200 AU, where the dust temperature is 16 K, would be $\sim$7000 yr, more than an order of magnitude longer than the desorption timescale.  In addition, the desorption time drops much more steeply with increasing dust temperature than the free-fall time, so even increasing $M_*$  by a factor 2 would not modify the desorption temperature by a perceptible amount. The light blue curve in Fig.  \ref{T_evap} shows  the free-fall time for a constant enclosed mass of 0.5 M$_\sun$, which is always  longer than the desorption time inside the respective snow line, independent of the assumed binding energy. We conclude that the infall motion of the envelope  will not have a significant impact on the location of the snow lines.

 \subsection{The N$_2$ binding energy}
 
 In order to match the observed ring-like emission of N$_2$ with our models, we had to assume an N$_2$ binding energy of 1000 K. This value is also higher than values measured in the laboratory for pure N$_2$ ices on N$_2$ surfaces using temperature programmed desorption. Again, the value of the binding energy depends on assumptions about the structure of the ices: if N$_2$ forms an ice layer above the CO ice, it will get sublimated earlier than if it is contained in a mixed CO-N$_2$ ice layer. In the latter case, the N$_2$ binding energy would be close to the CO binding energy. \citet{Oberg:2005} measured the binding energies of pure, layered, and mixed CO and N$_2$ ices. They find that pure N$_2$ ices have slightly lower binding energies than CO (790 $\pm$25 K versus 855 $\pm$ 25 K). In layered ices, the desorption kinetics are unchanged, but part of the N$_2$ desorbs together with the CO because N$_2$ diffuses into the CO ice. In a subsequent study, \citet{Bisshop:2006} specify the binding energy of pure N$_2$ ices as 800 $\pm$25 K. Our value of 1000 K agrees with the value measured for N$_2$ ices on a compact water ice surface with at least 0.4 monolayers (\citealt{Fayolle:2016}).

\citet{Oberg:2005} as well as \citet{Bisshop:2006} stress the important role of the ratio of the binding energies of N$_2$ and CO ices for the reproduction by chemical models of the observed anti-correlation. It is intuitively clear that this ratio needs to be smaller than 1, because otherwise all the N$_2$H$^+$ is subject to destruction by CO if it is released only when CO is also present in the gas phase. For pure ices \citet{Oberg:2005} and \citet{Bisshop:2006} find a ratio of $\sim$0.93, while for mixed ices the empirical value is lower at $\sim$0.89. The importance of this ratio for the observed anti-correlation indicates that a change in the CO binding energy as discussed in Section \ref{EbCO} may indeed imply an increase in the N$_2$ binding energy. The ratio of binding energies used in this study, determined by fitting the ring of N$_2$H$^+$ emission in IRAS4B, is 0.83, which is lower than the above cited ratio for pure ices, but close to that expected for mixed ices. 
 
 \subsection{The CH$_3$OH binding energy}\label{ch3oh}
 
Among our four sources, only the Gaussian uv-fits towards IRAS4A seems to reliably trace the size of the compact methanol emission because of the high signal-to-noise ratio, because IRAS4A is the only source where the peak of the emission is spatially resolved, and because the Gaussian fit to the methanol emission in the uv plane does not appear very much influenced by the outflow (The radially averaged uv data are shown in Appendix \ref{uvplots}.). The large uv-fitted size on IRAS4B  is clearly measuring the size of the resolved pedestal along the outflow, not that of the central peak. Also the uv-fit towards L1448C shows an apparent elongation along the outflow axis. In both sources, the peak of the methanol emission is not spatially resolved by our observations.

The fit of the methanol peak emission in IRAS4A in the uv plane yields a FWHM of 1.2$\pm$0.06$''$. This value corresponds to twice the size of the region where water emission occurs, as observed by \citet{Persson:2012} with a FWHM 0.61$\pm$0.12$''$. If both fits are reliable in tracing the sublimation regions without a significant influence of the outflow activity, this would hint at the (model-independent) interpretation that methanol has a lower binding energy than water. Indeed, laboratory studies find slightly lower values for the binding energy of pure methanol ices (e.g. \citealt{Collings:2004}: 5530 K for methanol bound to a water ice surface vs. 5700 K for pure water as well as water-methanol mixtures, which corresponds to dust temperatures of 120 K vs. 125 K in our sources).  Furthermore, methanol is expected to be also present in CO dominated ice when CO hydrogenation via H atoms on interstellar ice surfaces occurs (e.g. \citealt{Cuppen:2009}).

If we want to derive a specific methanol binding energy from our observations, we however need to assume a model-dependent source temperature profile.
If we take the temperature profiles of \citet{Kristensen:2012}, we find a freeze-out temperature of less than 70 K in IRAS4A based on our observations, which is much lower than the reported laboratory values. There are (at least) four different interpretations of this finding: 1. We see methanol co-desorbing from an ice environment with much lower binding energy than water ices  like, e.g., CO dominated ices (see above), 2. The adopted temperature profiles are wrong at these small radii and the methanol binding energy is actually higher than indicated by the  \citet{Kristensen:2012} profiles (see Sect. \ref{uncertainties}),  3. Non-thermal desorption at lower temperatures than the sublimation temperature may be efficient in releasing methanol into the gas phase, or 4. The influence of the outflow on the methanol emission is the dominating factor in determining the emission size and thus we cannot make statements about the methanol binding energy based on our observations. Because of these uncertainties, we refrained from properly fitting our observations with a modified methanol binding energy.

 \subsection{CO abundances}\label{COabundance}
 
Typical ISM CO abundances in molecular clouds range from $\sim$0.6--2 $\times$ 10$^{-4}$ (\citealt{Ripple:2013} and references therein). Assuming a $^{16}$O/$^{18}$O isotopic ratio of 500, this corresponds to a C$^{18}$O abundance of $\sim$1.2--4 $\times$ 10$^{-7}$. In our models, we find inner CO abundances that are about one order of magnitude smaller than these values (see Table \ref{modelparam}). 

 The CO abundances we find inside the CO snow line are lower than findings of most previous studies of protostars.
Modelling IRAM 30m observations of a sample of Class 0 protostars, \citet{Alonso:2010} find the lowest value they consider as model-input for the inner C$^{18}$O abundance, 10$^{-7}$, yielding the best fit of their observations. They however note that their fits would be even better with yet a lower value. \citet{Yildiz:2010} and \citet{Yildiz:2012} obtained observations of several transitions in C$^{18}$O up to (10--9) towards IRAS2A, IRAS4A, and IRAS4B with the Herschel telescope and several ground-based single-dish observatories. Fitting these observations with a chemical jump model located at 25 K, they find inner C$^{18}$O abundances of (1-3) $\times$ 10$^{-7}$. These abundances are an order of magnitudes higher than our values for the same sources. 

However, we note that their inner C$^{18}$O abundances are not very well constrained towards smaller values in their $\chi^2$ plots. Indeed, we have checked that our models for IRAS4A and 4B predict intensities consistent to those observed by \citet{Yildiz:2012} up to J = 10--9. Hence, it appears that spatially resolved emission profiles of C$^{18}$O such as those obtained here with the PdBI are essential to reliably constrain the gas-phase CO abundance inside the snow line. 
Using observations from the same group of observatories, \citet{Fuente:2012} derive a central C$^{18}$O abundance of 1.6 $\times$ 10$^{-8}$ towards the young intermediate-mass protostar NGC 7129 FIRS 2, which harbours a hot core.   

The low CO abundances are not restricted to the protostellar phase, they also seem to prevail in protoplanetary disks. \citet{Favre:2013} used observations of HD in the TW Hya disk to probe the H$_2$ mass in the warm gas at temperatures above 20 K. They find that the CO abundances in this warm molecular layer is reduced by at least an order of magnitude compared to dense clouds. They measure a disk-averaged warm gas-phase CO abundance relative to H$_2$ of (0.1-3) $\times$ 10$^{-5}$. Previous studies (e.g. \citealt{Dutrey:2003,Chapillon:2008,Qi:2011}) had already described the very low CO gas-phase abundance in disks, however these earlier studies were not able to differentiate between the warm and the cold layers and therefore could not exclude that the low value was due to depletion. 

There are several scenarios that might  explain these low CO abundances. 
One possible explanation could be that part of the CO is trapped in water ice and finally co-desorbs together with the water inside a region that is not resolved by our observations (see e.g. \citealt{Martin:2014} and Section \ref{EbCO}).
 Alternatively, CO in grain mantles could be removed by reactions to form other species (e.g. \citealt{Whittet:2011}). Observations confirm that a large part of carbon in the grain mantles can be bound in CO$_2$ ices that have a higher binding energy than CO (e.g. \citealt{Pontoppidan:2008}). Depending on the efficiency of gas-phase reactions that transform CO$_2$ into CO once the ice is sublimated to the gas phase, carbon could thus also be "hidden" in CO$_2$. Correlations of ice bands with A$_v$ show that the amount of CO and CO$_2$ in ices are comparable up to 25 mag and are each on average $4 \times 10^{-5}$ with respect to H$_2$ (eg. \citealt{Whittet:2007}). Our CO gas-phase abundances of $\sim$1--2.5 $\times$ 10$^{-5}$ are still a factor 2--3 times smaller than this. Hence CO ice would need to be more efficiently channeled into other species inside protostellar envelopes than in dark clouds.
 
By using a detailed gas-grain chemical model for protoplanetary disk conditions, \citet{Reboussin:2015} find that at high densities CO can indeed be efficiently converted in less volatile species like CO$_2$. Possible other candidate products are CH$_3$OH and CH$_4$. Indeed, observations indicate that methanol ice abundances in low-mass YSO envelopes can reach 15-25\% of water ice (\citealt{Pontoppidan:2008}), similar to the CO ice abundance in less dense clouds (\citealt{Whittet:2007,Whittet:2009}), while CH$_4$ reaches 5\% of the water ice (\citealt{Oberg:2008}). Some of the missing CO could also be in the form of more complex organic molecules (e.g. \citealt{Whittet:2011})
Two of the sources discussed in this paper, IRAS4A and IRAS4B, are well known hot corino sources that show a rich spectrum of COM emission lines. In order to properly investigate the relation between CO abundance and the existence of COMs, COM abundance measurements need to be performed for the sources of this study. In an upcoming paper (Belloche et al. in prep.), the CALYPSO project will study the COM emission in our sample of sources and might shed some light on their relevance for the problem of lacking carbon. However, the recent study by \citet{Taquet:2015} on IRAS4A already indicates that the abundances of COMs are not high enough to solve the problem. 

Finally, \citet{Aikawa:1997} suggest a chemical mechanism in protoplanetary disks where X-rays from the central star produce He$^+$ that can extract C from CO in the gas-phase. Part of the carbon is then transformed into hydrocarbons or CO$_2$. If X-rays are already produced in the Class 0 protostellar phase (which is not clear 
observationally, cf. \citealt{Prisinzano:2008}), this explanation could also be relevant for the presented sources. 

Given the small number of sources, it is obvious that more studies, particularly testing all the various scenarios  that may explain the lack of CO independently, are necessary in order to settle the question on where the missing carbon is. In any case, it seems that the low CO abundances that are found in protoplanetary disks are already established in the protostellar phase.

 \subsection{Comparison with  \citet{Jorgensen:2015}}
 
In a recent study, \citet{Jorgensen:2015} analysed the sublimation of CO, traced by emission in C$^{18}$O (2--1), for a sample of 16 protostellar sources in order to trace variations of protostellar accretion rates. Their work has three sources in common with ours: IRAS4A, IRAS4B, and L1448C. Their findings, however, deviate from ours both with respect to the sublimation radii determined by uv-fits of their observations and their modelling results. Their data was obtained with the SMA at an angular resolution of 2-3$''$, which is about a factor of three worse than ours. 

The elongation of CO emission along the outflow axis that we see in our data is not that apparent in their lower-resolution data. The difference is particularly striking in the case of IRAS4B. This fact may already cause differences in the results of the uv-fits, particularly as their observations can satisfactorily be fit by circular Gaussians while these are not at all well-suited to fit our observations. Their fits are however consistent with our fits perpendicular to the outflow, if we take their maximum uncertainties of 0.5$''$ and if we account for the fact that our elliptical fit of L1448C overestimates the extent of the emission in the image plane.

Regarding the applied modelling, there are clear differences between the two studies. In contrast to our modelling approach, they use a generic envelope model for all the sources. In particular, they apply a power law density profile in all cases with a fixed exponent of 1.5 and vary the source luminosity. The density profile is normalized by comparison to single-dish submillimeter continuum observations. Based on each luminosity, they determine a self-consistent temperature profile. In order to determine a best-fit model for each source, they subsequently compare their observed CO maps with the modelled CO emission extent. With this procedure, they can then compare the luminosity that yields their best-fit model with the current observed luminosity. If the current luminosity is much smaller than the modelled one, they count it as sign for a past accretion burst. As freeze-out temperature they take a value of 30 K. The luminosity values they use differ from ours for IRAS4A and L1448C: 9.9 versus 9.1 L$_{\sun}$ and 8.4 versus 9.0 L$_{\sun}$, respectively. For the three sources, their modelled emission regions are clearly smaller than ours. This difference is easily explained given their higher assumed freeze-out temperature.
 
Despite these differences, \citet{Jorgensen:2015} agree with our findings in their conclusion that the sources we have in common do not show signatures of a past accretion burst. We however note that our sample includes one source, namely IRAM 04191, that was not observed by \citet{Jorgensen:2015}  and that might have been subject to a past accretion burst. This source will be analysed in detail in a forthcoming paper.

 \section{Conclusions}\label{Conclusions}
 
As part of the CALYPSO Large Program, observations of C$^{18}$O, N$_2$H$^+$ and CH$_3$OH towards the nearest low-luminosity Class 0 protostars with the IRAM Plateau de Bure interferometer at (sub-)arcsecond resolution were obtained. For four sources of the CALYPSO sample, we analysed these observations to obtain information on the sublimation regions of different kinds of ices, which sheds light on the chemistry of the envelope, its temperature and density structure, and the history of the accretion process. This analysis comprised modelling the emission using the chemical model Astrochem coupled with the radiative transfer module RATRAN, using temperature and density profiles from \citet{Kristensen:2012}. 
 The main findings of our study are the following:
 
 \begin{enumerate}
 \item We observe an anti-correlation of C$^{18}$O and N$_2$H$^+$ in IRAS4B, IRAS4A, L1448C and L1157, with N$_2$H$^+$ forming a ring (perturbed by the outflow) around the centrally peaked C$^{18}$O  emission. In addition we observe compact methanol emission towards three of the sources.
\item The C$^{18}$O emission shows complex morphologies, influenced by the outflows. 
Gaussian uv-fits of the C$^{18}$O emission reveal the extent of the CO emission in these sources at radii (HWHM) of $\sim$280 to $\sim$560 au perpendicular to the outflow direction. 
\item Using a CO binding energy of E$_{\rm b}$=1200 K that corresponds to a dust temperature of $\sim$24 K in our models, and a N$_2$ binding energy of E$_{\rm b}$=1000 K that corresponds to a dust temperature of $\sim$19 K, we can satisfactory reproduce our observations towards all four sources. The modelling locates the CO snow lines at radii of 770 au, 730 au, 540 au, and  460 au for IRAS4A, L1448C, L1157, and IRAS4B, respectively. These values are substantially larger than the observed HWHM. The same holds for the relation between the radii of the CO snow lines and the HWHM derived from Gaussian uv-fits of the modelled C$^{18}$O emission, where the former are a factor of 1.5 -- 2 larger than the latter.
\item The value of E$_{\rm b}$(CO), is higher than the measured value for multi-layer desorption of pure CO (e.g. \citealt{Oberg:2005}). It however agrees with experimental values for monolayer desorption of CO from a water surface. This interpretation is also consistent with the low CO abundances we obtain in the warm gas in the central region of the envelope. Alternatively, the high binding energy could hint at CO being mixed with other ice components like e.g. CO$_2$ or organic molecules.  
\item We find very low CO abundances inside of the snowlines in the innermost part of the envelopes of our sources, more than an order of magnitude lower than the total CO abundance observed in the gas at large scales in molecular clouds. These values are comparable to CO abundances found in the warm molecular layer inside of protoplanetary disks. Possible reasons might be the conversion of CO ices into other species like CO$_2$ or, to a lesser extent, into complex organic molecules, the destruction of CO by X-rays, or partial entrapment of CO in water ices.  
\item In IRAS4A, the source where we spatially resolve the methanol emission with a high signal to noise, a Gaussian uv fit of the methanol emission yields a FWHM that is twice the size of the FWHM of water emission, observed by \citet{Persson:2012}. If both fits are reliable, this would hint at methanol having a clearly lower binding energy than water. We can, however, not rule out a strong influence of the outflow on the methanol emission, which might dominate over the effect of radiative heating of the envelope by the source.
\item We have also modelled the observed methanol emission. However, due to the large uncertainties of the applied source model in the innermost envelope and limitations of our observations with respect to signal-to-noise, angular resolution, and unresolved outflow contamination for this line, we cannot make precise quantitative statements on the methanol binding energy and abundances. 
 \end{enumerate} 
 
\begin{acknowledgements}
 We are grateful to an anonymous referee for useful comments
that helped to strengthen the paper. The research leading to these results has received funding from the French Agence Nationale de la Recherche (ANR), under reference ANR-12-JS05-0005. The project has received further support from the European Research Council under the 
European Union's Seventh Framework Programme (ERC Advanced Grant Agreement No. 291294 - `ORISTARS'). L.P. has received funding from the European Union Seventh Framework Programme (FP7/2007-2013) under grant agreement No. 267251.
We would also like to thank Bilal Ladjelate for his help in estimating the internal luminosities 
of the target sources from Herschel data.
 Furthermore, the authors would like to thank Edith Fayolle, St{\'e}phane Guilloteau, Lars Kristensen, Karin \"Oberg, and Magnus Persson for valuable discussions, comments, and suggestions. 
 \end{acknowledgements}

   \bibliographystyle{aa}
\bibliography{biblio}


\begin{appendix}

\section{The full sample of sources}\label{others}

Among the 16 sources that were observed in the CALYPSO programme, we chose the four sources that show a clear anti-correlation in their emission of C$^{18}$O and N$_2$H$^{+}$, for our analysis. In the remaining sources, we detect C$^{18}$O at a level of more than 3$\sigma$ in all cases, and in six sources at a level of more than 9$\sigma$ (L1527, IRAS2A, SVS-13A, L1448-2A, GF9-2, SERP-SMM4). In two cases, C$^{18}$O is detected towards two positions (SVS13, AQU-MMS1, SERP-SMM4) at a level of more than 6$\sigma$. N$_2$H$^+$ is detected in all fields of a size of 10$''$ around the main source at a level of more than 3$\sigma$. Three sources show only weak emission ($<$6$\sigma$: L1527, L1521F, AQU-MMS2), while the other sources show emission levels of 7-11$\sigma$.

In the sample, four different morphologies can be distinguished:
\begin{enumerate}
\item One-sided anticorrelation, where the source appears located at the edge of an N$_2$H$^+$ filament, strong emission in C$^{18}$O  and adjacent peak in N$_2$H$^+$ 
\item No or weak emission of N$_2$H$^+$, or no or weak emission of C$^{18}$O in our data
\item C$^{18}$O and N$_2$H$^+$ are both detected towards the source without a clear anti-correlation, relatively weak emission in C$^{18}$O, strong, relatively uniform emission in N$_2$H$^+$ 
\item Compact C$^{18}$O emission towards the source, no N$_2$H$^+$ emission towards the source, large N$_2$H$^+$ ring around the source not adjacent to the C$^{18}$O emission.
\end{enumerate}
 
According to this scheme, the observed sources could be classified as follows: 
\begin{enumerate}
\item IRAS2A, L1448-2A, SVS-13A, GF9-2, SERP-S68N, SERP-SMM4, AQU-MMS1 
\item L1527, L1521-F, AQU-MMS2
\item L1448-N, SVS13-B
\item IRAM04191
\end{enumerate}

\begin{figure}[!htb]
   \centering
   \includegraphics[width=0.5\textwidth]{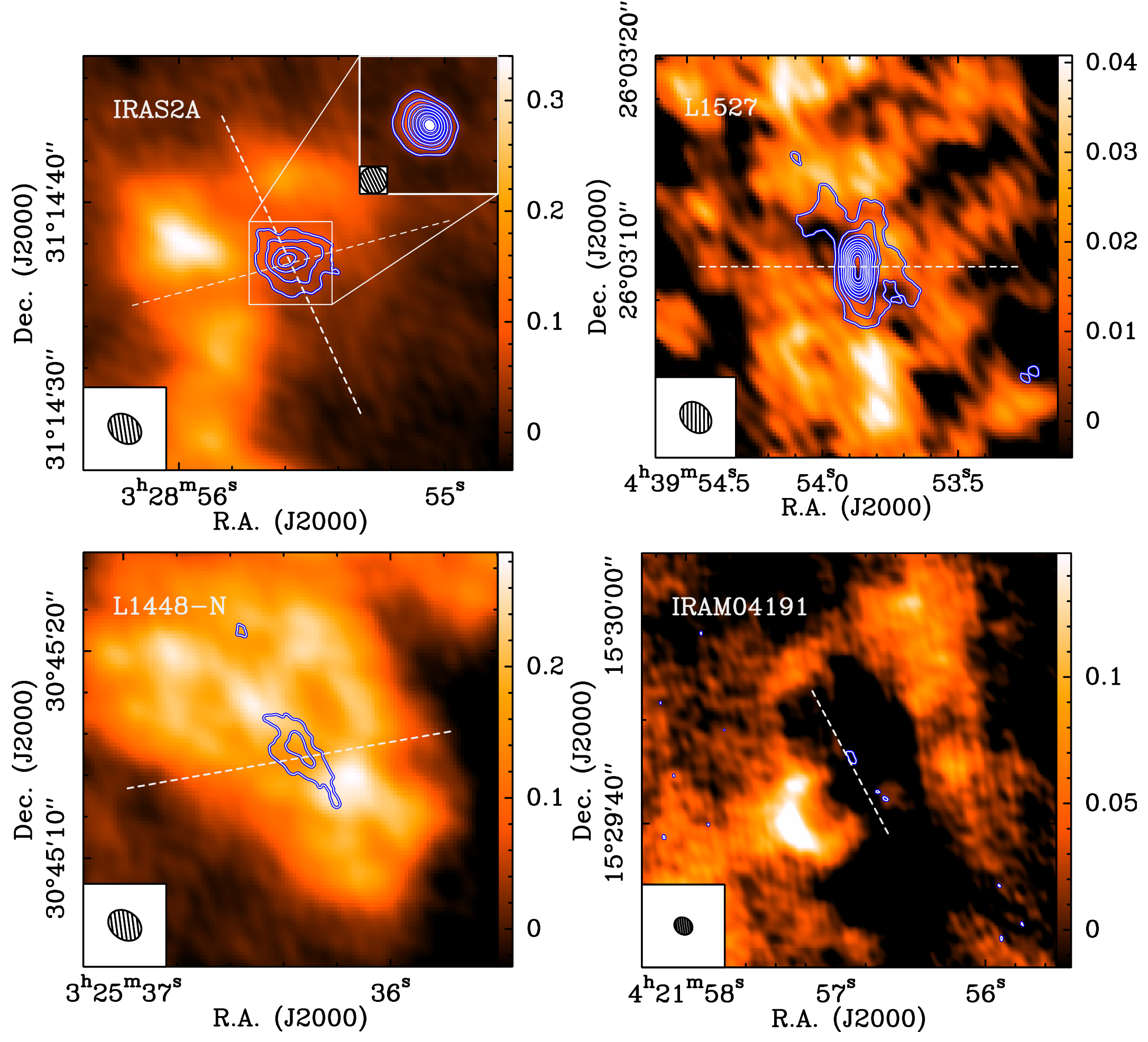}
      \caption{ N$_2$H$^+$ and C$^{18}$O integrated intensity maps for IRAS2A (top left), L1527 (top right), L1448-N (bottom left), and IRAM04191 (bottom right) as prototypical sources for the four morphology classes we find in the full sample of sources. Colour background: N$_2$H$^+$ (1--0) emission integrated over all seven hyperfine components. The noise in these maps is $\sigma$=(0.043, 0.012, 0.033, 0.032) Jy beam$^{-1}$ km s$^{-1}$ for IRAS2A, L1527, L1448-N, and IRAM04191, respectively. The wedges show the N$_2$H$^+$ intensity scale in Jy beam$^{-1}$ km s$^{-1}$. Note that while for the first three sources the FOV is 20$''$ $\times$ 20$''$, it is 40$''$ $\times$ 40$''$ for IRAM04191 in order to show the full ring-like morphology.  Contours show integrated emission  of C$^{18}$O (2--1) in steps of 3$\sigma$, starting at 3$\sigma$ up to 60$\sigma$, with $\sigma$=(0.037, 0.022, 0.038, 0.018) Jy beam$^{-1}$ km s$^{-1}$ for IRAS2A, L1527, L1448-N, and IRAM04191, respectively. The C$^{18}$O emission was integrated over $\pm$ 3 km s$^{-1}$ around the systemic velocity of each source. The inlay in the upper right corners of the top left panel shows the methanol emission towards IRAS2A as colour background and contours in steps of 3$\sigma$, starting at 3$\sigma$ up to 60$\sigma$ inside the central 2$''$. The filled ellipses in the lower left corner of the panels indicate the synthesized beam sizes of the N$_2$H$^+$ observations at 3 mm. The dashed white lines illustrate the small-scale outflow directions (cf. \citealt{Codella:2014}, \citealt{Hogerheijde:1998}, Podio et al. in prep., \citealt{Belloche:2002}). }
               \label{maps_A1}
   \end{figure}

We show four sources representing each of the different morphologies in Fig. \ref{maps_A1}: IRAS2A representing class 1, L1527 representing class 2, L1448-N representing class 3, and IRAM04191 representing class 4. The different morphologies seem to suggest that if a) the source emits strongly in C$^{18}$O and if b) the source is fully embedded in an N$_2$H$^+$-rich environment, an anti-correlation is always seen. If the source emits strongly in C$^{18}$O and sits at the edge of an N$_2$H$^+$ filament, the anti-correlation appears only one-sided (case 1). On the other hand, if there is either no C$^{18}$O or no N$_2$H$^+$, obviously no anti-correlation is observed at all (case 2 ). There also seems to be cases where C$^{18}$O does not appear abundant enough to significantly destroy N$_2$H$^+$ towards the source (case 3). As already mentioned in Sect. \ref{othersources}, IRAM04191 stands out in the sense that the large ring emission of N$_2$H$^+$ might hint at a past accretion burst (case 4).

   \section{CO and CH$_3$OH uv data}\label{uvplots}
   
\begin{figure}[!htb]
   \centering
   \includegraphics[width=0.43\textwidth]{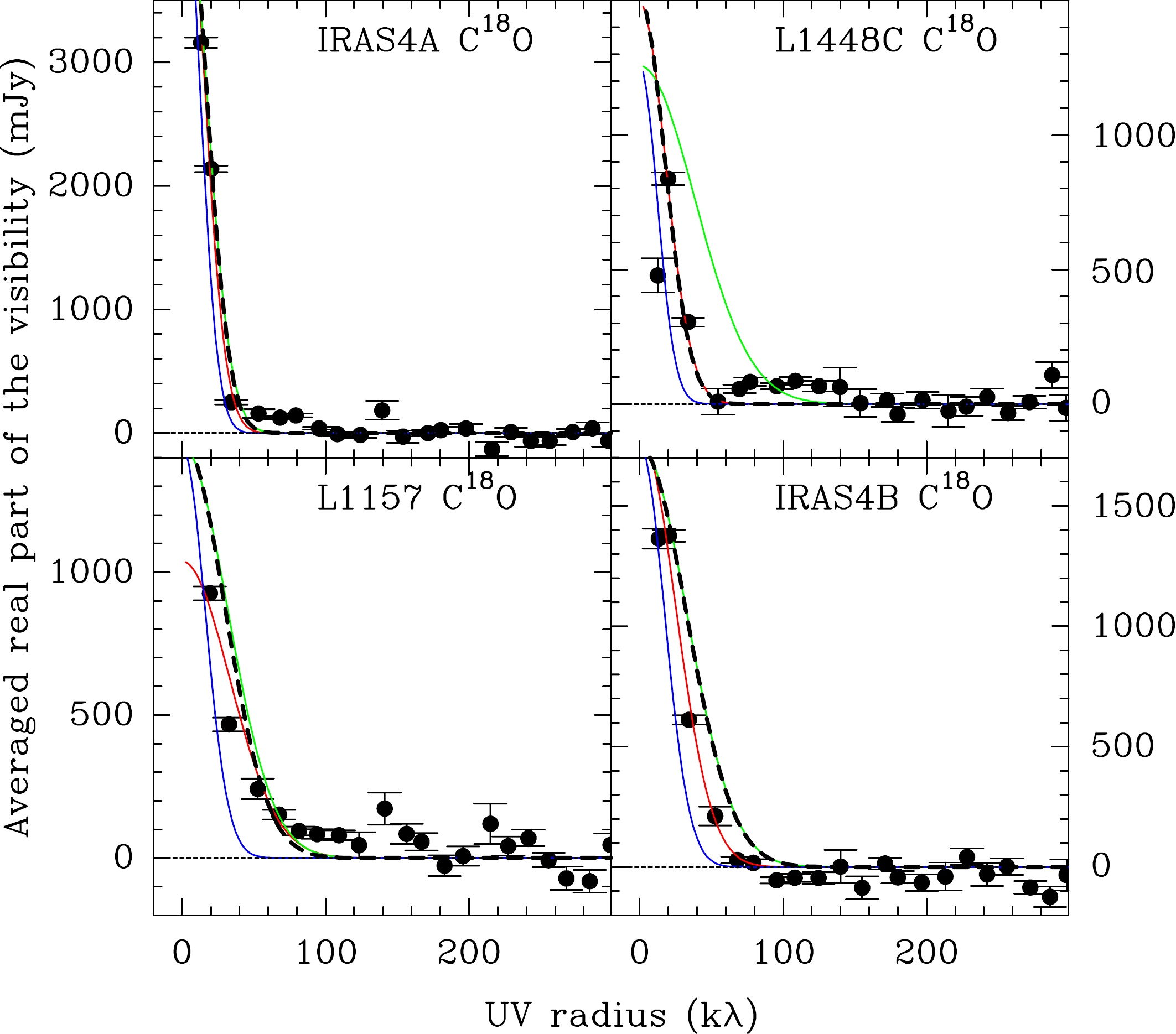}
      \caption{ C$^{18}$O uv-plots for IRAS4A (top left), L1448C (top right), L1157 (bottom left) and IRAS4B (bottom right). The data is integrated over $\pm$3 km s$^{-1}$ around the line centre and radially averaged in bins of 20 m with task \texttt{uv\_circle} in \texttt{MAPPING}. Red lines show the result of the circular Gaussian fits to the un-averaged uv data, while the blue and green lines show the results of the elliptical Gaussian fits to the un-averaged uv data along the major and minor axes, respectively. Black dashed lines show the result of the elliptical Gaussian fits with fixed P.A. along the outflow. The parameters of the fits are listed in Table \ref{fitsc18o}.}
               \label{uv_c18o}
   \end{figure}

\begin{figure}[!htb]
   \centering
   \includegraphics[width=0.43\textwidth]{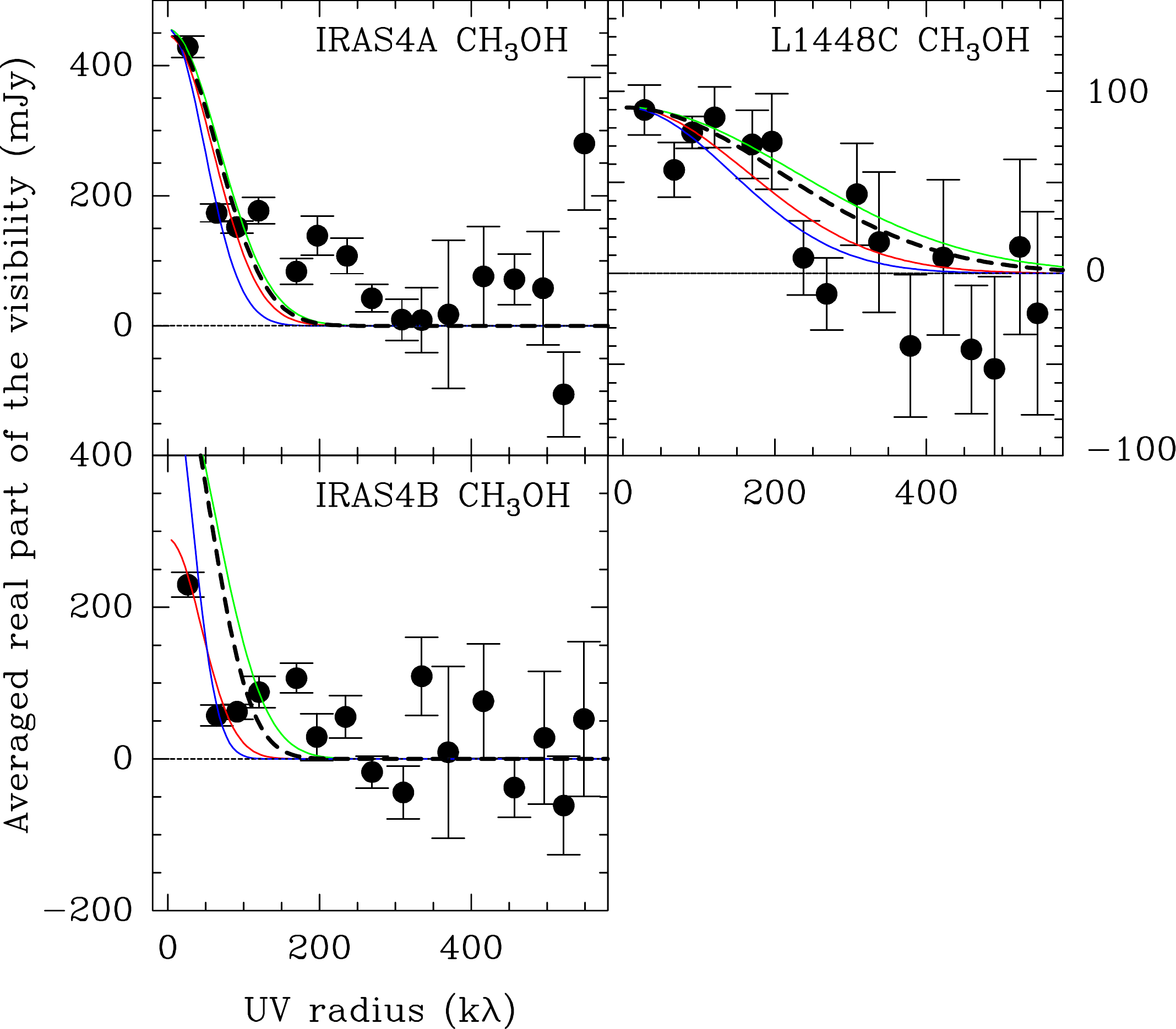}
      \caption{ Same as Figure \ref{uv_c18o}, but for CH$_3$OH in IRAS4A (top left), L1448C (top right), and IRAS4B (bottom left). The data is radially averaged in bins of 50 m with task \texttt{uv\_circle} in \texttt{MAPPING}. The parameters of the fits are listed in Table \ref{fitsch3oh}.}
               \label{uv_ch3oh}
   \end{figure}

In Section \ref{snowlines1}, we followed the standard procedure of using circular and elliptical Gaussian uv-fits in order to measure the size of the source emission in C$^{18}$O and CH$_3$OH. Figures \ref{uv_c18o} and \ref{uv_ch3oh} show plots of the uv data, obtained by radially averaging the data that was used for the fits listed in Tables \ref{fitsc18o} and \ref{fitsch3oh}. While the radial average hides all the complex spatial morphological structure that is visible in the maps shown in Fig. \ref{maps}, it may still give an impression of the quality of the data. For comparison, the results of the elliptical and circular Gaussian fits are shown as well, which may hint at the strength of radial variations mirrored in the eccentricity of the elliptical fits.

\section{Model parameter dependences}\label{chem}

\begin{figure*}[!htb]
   \includegraphics[width=0.9\textwidth]{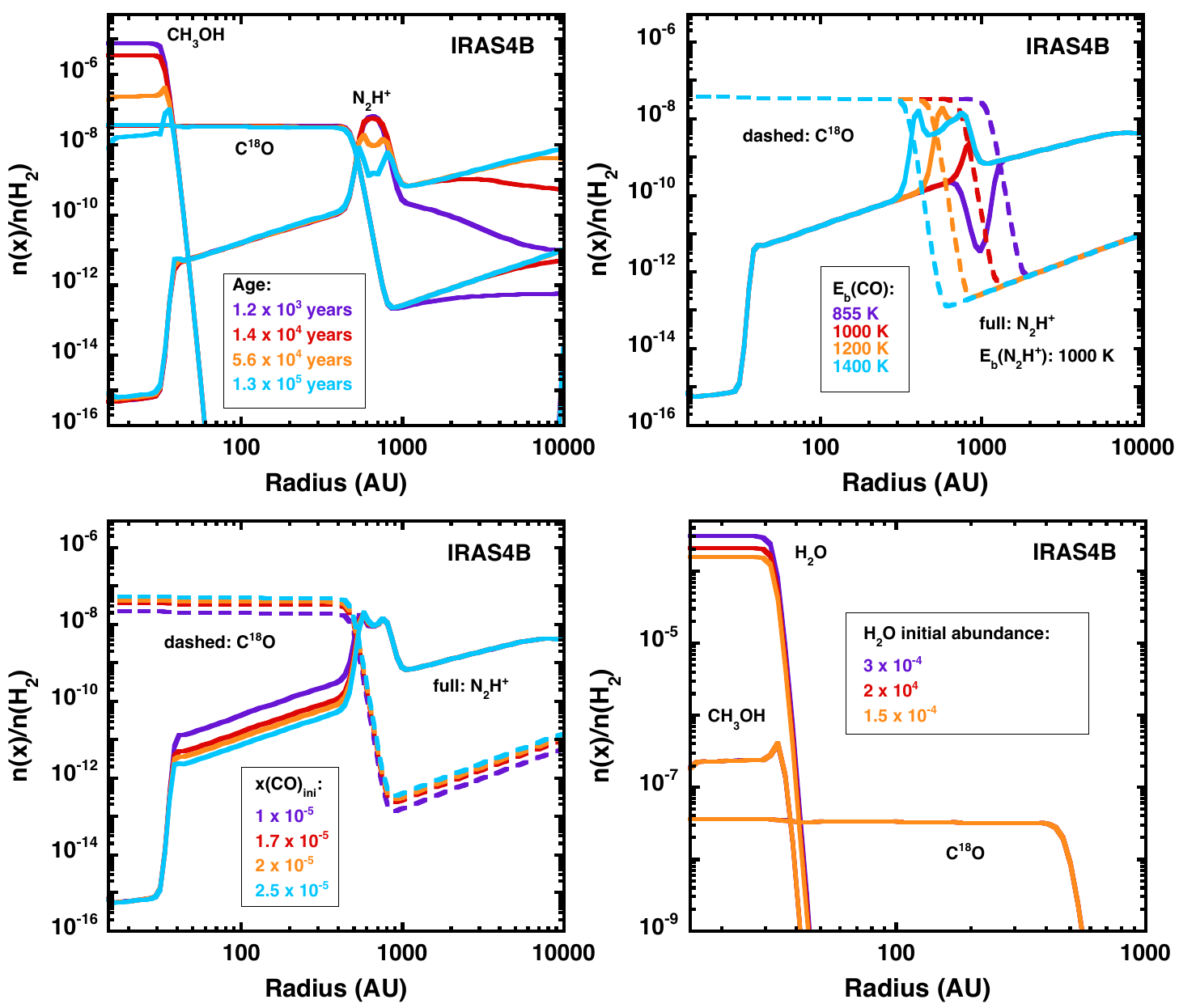}
      \caption{ Dependences of the chemical model on various free parameters. Top left: The fractional abundances of CH$_3$OH, C$^{18}$O, and N$_2$H$^+$ are shown for four different chemical ages. Same colours identify abundance curves at the same chemical time for chemical ages as indicated in the legend. Top right: The fractional abundances of C$^{18}$O (dashed lines) and N$_2$H$^+$ (full lines) are displayed for different values of the CO binding energy. Same colours identify abundance curves computed with the same CO binding energy as indicated in the legend. Bottom left: The fractional abundances of C$^{18}$O (dashed lines) and N$_2$H$^+$ (full lines) are displayed for different values of the initial fractional CO abundance. Same colours identify abundance curves computed with the same CO initial abundance as indicated in the legend. Bottom right: The fractional abundances of H$_2$O, CH$_3$OH, and C$^{18}$O are shown for four different values of the initial fractional H$_2$O abundance. Same colours identify abundance curves computed with the same H$_2$O initial abundance as indicated in the legend.    }
\label{chemistry_dependence}
   \end{figure*}

Figure \ref{chemistry_dependence} illustrates the dependence of the chemical model on the most important free parameters. As discussed in Section \ref{model}, the computed C$^{18}$O abundance is independent of the assumed chemical age, while CH$_3$OH and N$_2$H$^+$ get destroyed with time (see top left panel). Therefore, our observations do not constrain the initial fractional abundances for these latter two species, because the inference from the final abundances fed into Ratran to the initial abundances of CH$_3$OH and N$_2$H$^+$ relies on the unknown chemical age. The top right panel of Figure \ref{chemistry_dependence} shows how the CO emission size varies with the assumed CO binding energy. This plot also demonstrates that for the anti-correlation between CO and  N$_2$H$^+$ to be established, the N$_2$ binding energy has to be smaller than the binding energy of CO. The bottom left panel shows that all C$^{18}$O that is initially put into ices will be found in the gas phase inside of the snow line, such that the resulting CO peak intensity can directly be fine-tuned by varying the initial fractional CO abundance. Finally, as already discussed in Section \ref{model}, the bottom right panel demonstrates that both the CO and CH$_3$OH abundance profiles do not depend on the assumed initial water abundance.

\section{Initial abundances for Astrochem}\label{iniabu}

Table \ref{iniabun} lists the initial fractional abundances relative to H$_{2}$ used in the chemical computations with Astrochem. For the metals, the low-metal case as described in \citet{Wakelam:2008} was adopted. The molecules were initially put into ices, because the sublimation timescale is much smaller than the desorption timescale inside of the snow line (see Fig. \ref{T_evap}) such that the molecules return quickly into the gas phase in that region.

 \begin{table*}[!htb]
\footnotesize
\caption{Initial fractional abundances used in Astrochem, relative to H$_{2}$.}            
\label{iniabun}      
\centering                          
\begin{tabular}{l l l l l l }     
\hline\hline
\noalign{\smallskip}
Species   & Frac. abun. & Frac. abun. & Frac. abun. &  Frac. abun.& Remark/Reference\\
   & IRAS4B & IRAS4A & L1448C &  L1157&\\
\noalign{\smallskip}
 \hline           
\noalign{\smallskip}
H$_2$ & 1.00$\times$10$^{0}$& 1.00$\times$10$^{0}$ & 1.00$\times$10$^{0}$ & 1.00$\times$10$^{0}$&\\ 
He & 1.70$\times$10$^{-1}$ & 1.70$\times$10$^{-1}$ & 1.70$\times$10$^{-1}$ & 1.70$\times$10$^{-1}$ &\citet{Asplund:2009} \\
S$^+$& 1.60$\times$10$^{-7}$ & 1.60$\times$10$^{-7}$ & 1.60$\times$10$^{-7}$ & 1.60$\times$10$^{-7}$&\citet{Wakelam:2008}\\
Si$^+$&1.60$\times$10$^{-8}$ &1.60$\times$10$^{-8}$ &1.60$\times$10$^{-8}$ &1.60$\times$10$^{-8}$&\citet{Wakelam:2008}\\
Fe$^+$&6.00$\times$10$^{-9}$ &6.00$\times$10$^{-9}$ &6.00$\times$10$^{-9}$ &6.00$\times$10$^{-9}$&\citet{Wakelam:2008}\\
Na$^+$&4.00$\times$10$^{-9}$ &4.00$\times$10$^{-9}$ &4.00$\times$10$^{-9}$ &4.00$\times$10$^{-9}$&\citet{Wakelam:2008}\\
Mg$^+$&1.40$\times$10$^{-8}$ &1.40$\times$10$^{-8}$ &1.40$\times$10$^{-8}$ &1.40$\times$10$^{-8}$&\citet{Wakelam:2008}\\
e$^-$&2.00$\times$10$^{-7}$ &2.00$\times$10$^{-7}$ &2.00$\times$10$^{-7}$ &2.00$\times$10$^{-7}$ &\\
NH$_3$(ice)&4.28$\times$10$^{-6}$&4.28$\times$10$^{-6}$ &4.28$\times$10$^{-6}$ &4.28$\times$10$^{-6}$& 10\% of all N, \citet{Wakelam:2008}\\
CO(ice) & 1.70$\times$10$^{-5}$& 8.80$\times$10$^{-6}$&9.60$\times$10$^{-6}$& 2.80$\times$10$^{-5}$& free parameter to match observed intensity\\
N$_2$(ice)&4.80$\times$10$^{-7}$&3.00$\times$10$^{-7}$&4.60$\times$10$^{-7}$& 9.40$\times$10$^{-7}$& free parameter to match observed intensity\\
CH$_3$OH(ice)& 1.00$\times$10$^{-5}$& 3.60$\times$10$^{-6}$& 5.00$\times$10$^{-6}$&<2.00$\times$10$^{-6}$& free parameter to match observed intensity\\
H$_2$O(ice)&2.96$\times$10$^{-4}$&3.08$\times$10$^{-4}$&3.10$\times$10$^{-4}$&2.96$\times$10$^{-4}$&all remaining O, \citet{Hincelin:2011}\\
\noalign{\smallskip}
 \hline           
\end{tabular}
\end{table*}

\end{appendix}

 \end{document}